# 3-dimensional plasmonic nanomotors enabled by independent integration of Optical Pulling and Lateral Forces

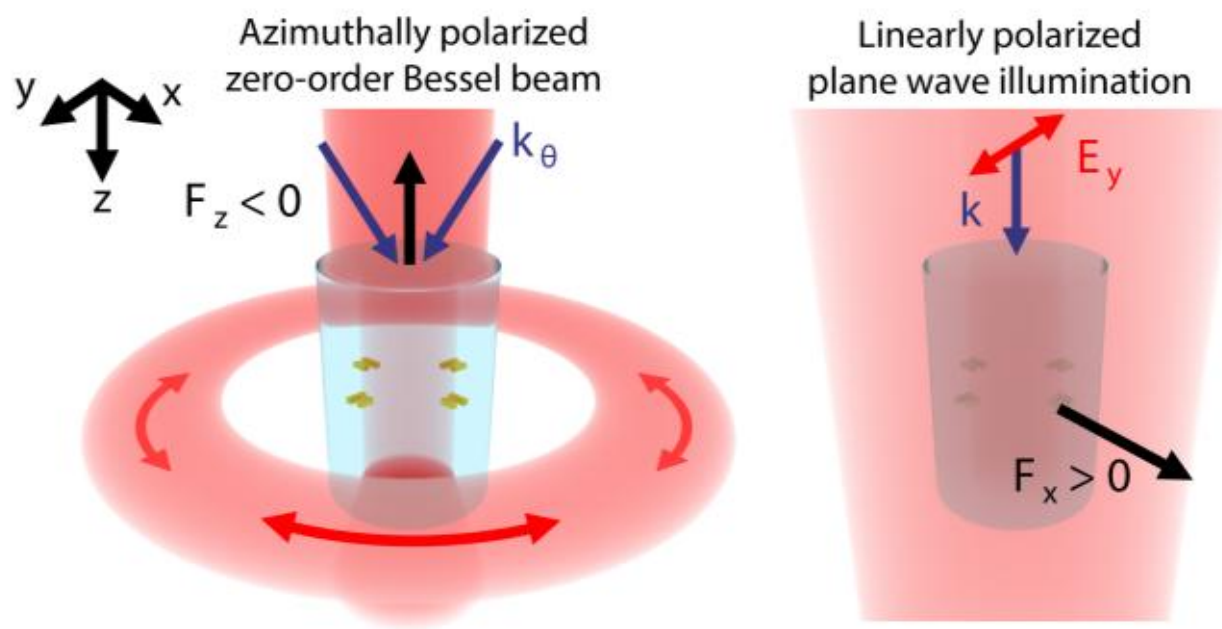

Guillermo Serrera[1], Yoshito Y. Tanaka[2] and Pablo Albella[1*]

*1. Group of Optics, Department of Applied Physics, University of Cantabria, 39005, Spain.*
*2. Research Institute for Electronic Science, Hokkaido University, N21W10, Kita, Sapporo, Hokkaido 001-0021, Japan.*
**pablo.albella@unican.es*

**Abstract:** : Light-matter interactions generally involve momentum exchange between incident photons and the target object giving rise to optical forces and torques. While typically weak, they become significant at the nanoscale, driving intense research interest in the exploitation of photon recoil to drive micro- and nanostructures. While great progress has been attained in controlling transversal degrees of freedom, three-dimensional movement remains challenging, particularly due to the impractical realization of pulling forces that oppose the direction of incident light. Here we theoretically present a novel nanomotor design that enables independent control over both transverse and longitudinal motion. This design exploits coupling between an azimuthally polarized Bessel beam and a dielectric glass cylinder to realistically achieve optical pulling forces. At the same time, asymmetric plasmonic dimers, embedded within the cylinder, provide lateral motion, through asymmetric scattering under plane wave illumination. We further demonstrate that unwanted displacements and rotations can be restrained, even at long illumination times. Our design unlocks a new degree of freedom in motion control, allowing for pulling, pushing, and lateral movement by simply tuning the polarization or switching between plane waves and Bessel beams.

## 1. Introduction

At the core of light-matter interaction, photons are scattered or absorbed by objects leading to an exchange or transfer of linear and angular momentum. This, by Newton's Third Law, results in optical forces and torques acting back on the interacting objects [1], [2]. The ability to remotely manipulate objects has prompted intense research into microscopic light-driven actuators, looking for the miniaturization and simplification of current systems, with a focus on lab-on-a-chip technologies [3], [4]. These actuators often include fluid pumps [5], [6], valves [7], and mixers [8], as well as motion actuators fueled by strong gradient forces in tightly focused beams [9], [10]. However, reliance on gradient forces imposes some design constraints, including the need for beam steering, its short working range due to the low Depth of Focus of high NA lenses, and the limitations imposed by diffraction into the light intensity gradient. Most importantly, the interplay between gradient and scattering forces enforces severe limitations on the actuated object's geometry and material.

All these constraints have motivated the study of scattering forces with plane waves or weakly focused beams as an alternative to optical tweezers and gradient-driven forces. In particular, the exploitation of asymmetric scattering from nanostructures has led to some successful lateral optical nanomotors. These schemes have relied on asymmetric plasmonic assemblies, such as gammadion shapes [11], dimers/trimers resembling Yagi-Uda antennas [12], [13], or dielectric metasurfaces with directional scattering [14]. Combinations of these elements allow for control of different degrees of freedom within the transversal plane, including both translation and rotation.

However, controlling longitudinal motions without gradient forces remains a challenge in this type of devices. 'Pushing' forces are a trivial byproduct of light scattering at normal incidences, arising as a direct result of momentum conservation. On the other hand, 'pulling' scattering forces are counter-intuitive, since $F_{long} = \frac{W_{scat}}{c}(\cos\theta_0 - \langle\cos\theta\rangle)$, where the longitudinal force $F_{long}$ is given by the scattered power $W_{scat}$ and the relationship between the cosine of the incident angle $\cos\theta_0$ and the scattering angular average $\langle\cos\theta\rangle$. This means that the optical pulling effect is attained when the object collimates the incident light ($\langle\theta\rangle < \theta_0$) [15]. Since normally incident plane waves are characterized by $\theta_0 = 0$, they cannot sustain optical pulling forces. As a result, alternative methods must be considered, like the use of internal bubbles [16], topology-engineered materials [17], [18], interactions with interfaces [19], strong chiral light-matter interactions [20], [21], exploiting symmetry in Bound States in the Continuum [22] or complex optical conveyors [23], [24], [25].

Nonetheless, the two most successful approaches for realizing optical pulling forces have been the use of Bessel beams [26] and Optical Tractor Beams (OTBs), based on the interference of several obliquely incident plane waves [27], [28]. In both cases, the individual photons exhibit oblique incidence angles $\theta_0$. While pulling forces have been successfully achieved using very large angle Bessel beams and OTBs, they are impractical due to their short range [29]. In this respect, azimuthally polarized Bessel beams, coupled to anti-reflective coated dielectric cylindrical waveguides have shown to be an effective approach to reduce angles below 45º, where range is no longer suppressed [30].

While these various techniques for optical pulling exist, their realistic application in the control of longitudinal motion in nanomotors is still missing, and current designs, based on sophisticated responses to different incident polarizations, are not suitable for this purpose.

In this work, we propose a realistic nanomotor design whose motion can be independently controlled in both the transversal plane and the longitudinal axis. A suitable platform for optical pulling was first developed, based on waveguide coupling between an azimuthally polarized Bessel beam and a glass cylinder coated with anti-reflective layers. Then, using this cylinder as the main chassis for the nanomotor, asymmetric plasmonic dimers were strategically placed within the dielectric cylinder to avoid strong interaction with the pulling beam and thus preserving the optical pulling effect. Therefore, a plane wave incidence produces lateral motion, driven by the lateral force from the plasmonic dimers, while incidence with a Bessel beam provides movement on the longitudinal axis via the optical pulling force on the cylinder. Our design enables the selection of motion modes: pulling, pushing, and lateral movements (left-right, forward-backward) by adjusting the polarization or switching between plane wave and Bessel beam illuminations. Furthermore, we demonstrate that the optical pulling force achieved in our system is highly robust against rotations, translations and Brownian motion, furthering the versatility and practical applicability of our device.

## 2. Conceptual overview and optical pulling platform

As showcased by authors in [30], optical pulling forces (OPFs) benefit from transversely isotropic particles (like cylinders) and beams, with the azimuthal polarization being more efficient in the compromise between pulling force and Bessel beam cone angle. These beams are characterized by an electric field

$$\boldsymbol{E}(\rho,\varphi,z) = -iE_0 e^{ik_0\cos\theta_0 z}\,\frac{J_0'(k_0\sin\theta_0\,\rho)}{\sin\theta_0}\,e^{im\varphi}\,\boldsymbol{n_\varphi} \tag{1}$$

where $E_0$ is the amplitude, $\mathrm{k}_0$ is the beam wavenumber, $J_0'(x)$ represents the first derivative of the Bessel function $J_0(x)$ and $m$ is the topological charge of the beam, which will be zero hereinafter for the rest of this work. Upon interaction with these beams, dielectric cylinders with appropriate diameters (see Supplementary Information) can support Fundamental Waveguide Modes (FWMs), helping reduce photon recoil in the structure. This effect can be enhanced with the usage of antireflection coatings (ARCs). As stated before, OPFs need collimation of the incident light, something that can happen for some values of the length of the cylindrical waveguide, owing to interference of some Fourier components of the scattered field.

The general idea behind our design is illustrated in **Figure 1a-b**. We exploit the efficient OPF from zero-order azimuthally polarized Bessel beams on dielectric cylinders, and we add plasmonic rod dimers to provide lateral motion. As seen in **Figure 1c**, these plasmonic dimers, when illuminated with $y$-axis polarization at the superposition of their respective resonance (around 980 nm in **Figure 1c**), provide very directional lateral scattering, driving the lateral ($x$-component) force seen in **Figure 1d** [31], [32], [33]. Meanwhile, the perpendicular polarization at the same wavelength, is characterized by very low and uniform scattering, resulting in negligible optical forces, as shown in **Figure 1d**.

As can be seen in **Figure 1d**, the optical force exerted in these dimers has a dominant longitudinal ($z$) component, which closely follows the dimer's extinction cross-section. This $z$-component force is also quite large for even moderate illumination intensities (0.4 mW $\mu m^{-2}$), meaning that the pulling force exerted by a Bessel beam might be suppressed if strong interaction with the dimers occurs. This interaction can be avoided by employing wavelengths higher than 1000 nm and a polarization perpendicular to the dimer's orientation, so that no resonance is excited. To achieve a cylinder design working at such wavelengths, we explore different combinations of Bessel beams (1 mW $\mu m^{-2}$ maximum intensity) with wavelengths $\lambda \in [1000,1500]$ nm and refractive indices for the dielectric cylinder $n_p \in [1.6,1.8]$. For each of these combinations, we investigate Bessel beam angles $\theta_0 \in [10,50]$°, each of them corresponding to different cylinder diameters; and different cylinder lengths $L \in [0.5, 3]$ µm. Finally, each of the cylinders is equipped with suitable ARCs (see Supplementary Information). The background refractive index is set to that of water $n_b = 1.33$.

The dynamics of a cylinder in an aqueous medium strongly depend on its geometrical aspect ratio, defined as $p = L/D$, where $L$ is the cylinder's length and $D$ its diameter. To investigate such different dynamics, we select two suitable configurations characterized by different cylinder length/diameter aspect ratios, as shown in **Figure 1e-f**. In particular, **Figure 1e** shows the longitudinal ($z$-component) force map for a 1500 nm Bessel beam and a 1.6 index cylinder, where long cylinders are generally required to excite the FWM and generate the pulling force, with a maximum negative force of -1.2 pN, located at an angle of 35º (1.77 µm diameter) and a length of 3 µm (2.04 aspect ratio).

On the other hand, **Figure 1f** shows the force map for a 1300 nm wavelength Bessel beam and a 1.7 index cylinder, where the maximum negative force of -0.6 pN is located at an angle of 40º (1.44 µm diameter) and a length of 1.8 µm. To ensure that we can work with a sufficiently different aspect ratio, here we select a suboptimal configuration (same diameter, length of 1.5 µm) with a lower (1.44) length/diameter aspect ratio.

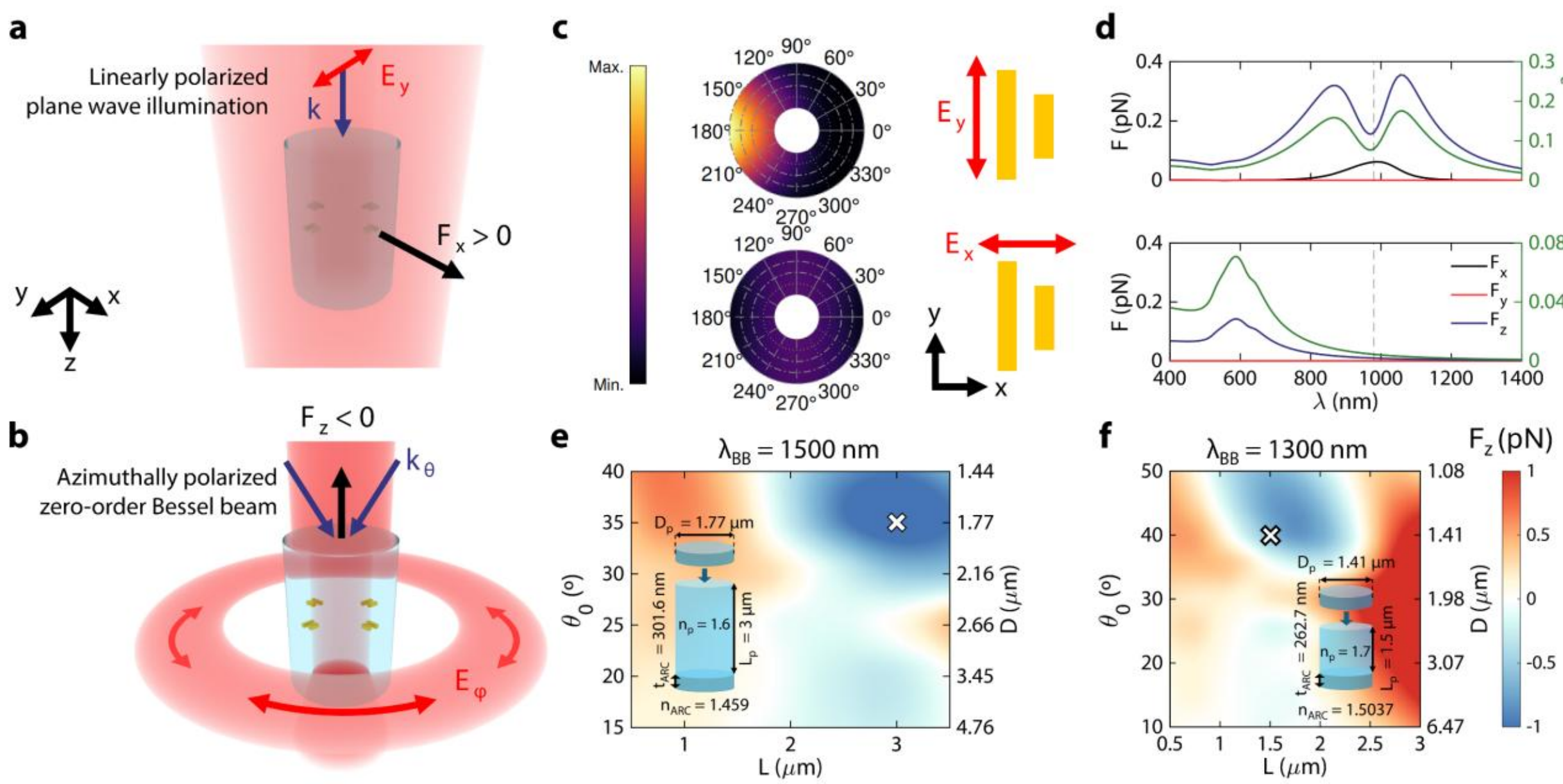

**Figure 1.** Scheme of the nanomotor's lateral movement, provided by the plane wave illumination (a) and its pulling force mechanism with the Bessel beam (b). c) Scattering pattern of plasmonic nanorod dimers at a wavelength of 980 nm, for both parallel ($E_y$) and perpendicular polarizations ($E_x$). The nanorods are characterized by lengths of 170 and 130 nm, separated by a distance of 100 nm. Their depth and width are the same, at 50 nm. d) Optical forces and extinction cross section for the aforementioned dimers and polarizations. The illumination intensity is 0.4 mW µm$^{-2}$. e) Longitudinal force on a dielectric cylinder upon interaction with an azimuthally polarized Bessel beam (characterized by angle $\theta_0$) at a wavelength of 1500 nm and a wavelength of 1300 nm (f). The insets reflect the geometrical and material features of the two selected designs, corresponding to the white crosses in their respective maps.

## 3. Plasmonic dimer placement for optical pulling preservation and lateral motion

As stated previously, minimizing the interaction with the dimers is key to preserve the pulling force. As shown in **Figure 1d**, interaction is minimized for polarizations orthogonal to the dimers, given that the wavelength is sufficiently long as to avoid the resonance of the rod's short dimension and the intrinsic absorption of gold. However, the complex interaction of the Bessel beam with the nanomotor's glass chassis might result in complex polarization patterns, meaning that the force exerted on the dimers can be difficult to predict.

To characterize the general properties of the incident field impinging on the dimers, we show in **Figure 2a** the electric field inside and in the vicinity of the cylinder in **Figure 1f** (1.44 µm diameter and 1.5 µm length), upon interaction with its suitable pulling Bessel beam. Cut planes below, in the middle and above the cylinder are featured in **Figure 2b-d**, with the field's polarization. As can be seen in **Figure 2a**, the field has a similar shape to the incident beam below the cylinder, with a clear azimuthal polarization pattern in **Figure 2d**. Meanwhile, in the inner part of the cylinder field gets concentrated around its center due to its higher refractive index. Due to the TE FWM excitation, the field retains its azimuthal polarization (**Figure 2c**). Finally, the field above the cylinder has a more complex shape, owing to interference with the parts of the beam that impinge laterally on the cylinder. However, a close examination in **Figure 2b** shows that the field is generally less focused due to the collimation effect and remains azimuthally polarized.

In general, the incident field on the dimers will correspond to the same profile of the incident Bessel beam, corrected by the higher refractive index of the cylinder. Thus, we can find the Lorentz's force exerted on a single nanorod by approximating said rod as a dipole with dipolar moment $\boldsymbol{p}$. For harmonic fields, the time-averaged force $\langle \boldsymbol{F} \rangle$ is [34], [35]:

$$\langle \boldsymbol{F} \rangle = \frac{1}{2} \Re\mathrm{e}[(\boldsymbol{p}^* \cdot \nabla)\boldsymbol{E} + i\omega(\boldsymbol{p}^* \times \boldsymbol{B})] \tag{2}$$

where $\omega$ is the angular frequency of light and $\boldsymbol{B}$ is its magnetic field. Using the electric field from **equation 1** with $m = 0$ and letting $\boldsymbol{p} = \varepsilon_b \alpha \boldsymbol{E}$ (with $\varepsilon_b = \varepsilon_0 n_b^2$) and, from Maxwell's equations, $\boldsymbol{B} = (i\omega)^{-1} \nabla \times \boldsymbol{E}$; it can be shown (see Supplementary Information for the full derivation) that the total force is

$$\langle \boldsymbol{F} \rangle = \frac{1}{2} \Re\mathrm{e}\left[\varepsilon_b \alpha^* E_\varphi^* \left(\frac{\partial E_\varphi}{\partial \rho} \boldsymbol{n_\rho} + \frac{1}{\rho} \frac{\partial(\rho E_\varphi)}{\partial z} \boldsymbol{n_z}\right)\right] \tag{3}$$

where $E_\varphi$ is the azimuthal field in **equation 1.** This expression can be further simplified to

$$\langle \boldsymbol{F} \rangle = \frac{1}{2} \Re\mathrm{e}\left[\varepsilon_b \alpha^* E_0^2 \left(\frac{k_0}{\sin\theta_0} J_1(x_\rho) J_1'(x_\rho) \boldsymbol{n_\rho} + i \frac{k_0 \cos\theta_0}{\sin^2\theta_0} J_1^2(x_\rho) \boldsymbol{n_z}\right)\right] \tag{4}$$

where $x_\rho = k_0 \sin\theta_0 \, \rho$. As can be seen, a transversal force in the radial direction arises as well as a longitudinal force. Note that the conjugate polarizability $\alpha^*$, at the Frölich condition of a plasmonic nanoparticle, will have a negative imaginary part [36], meaning that the longitudinal force will be positive (pushing force). Both force components can be eliminated by: (1) placing the dipole at zero field points (zeroes of the Bessel function $J_1(x_\rho)$), which can be impractical due to the size constraints imposed by the optical pulling force as well as fabrication issues; and (2) minimizing the polarizability, making both force components effectively zero and allowing independent control of pulling and lateral forces.

For rod-like particles, characterized by highly anisotropic polarizabilities, interaction can be suppressed by polarizing light along the shorter dimension of the rod, minimizing and shifting the resonance to lower wavelengths. Although interactions in a dimer can result in more intricate force patterns, a parallel rod dimer, as previously shown in **Figure 1d**, can present a similar response to a single rod system. This means that parallel rod dimers, in contrast with other proposals, could be employed to effectively select between transversal and longitudinal motion.

Due to the non-trivial interactions between the rods in a dimer, we propose a semi-analytical Coupled Dipole Approximation (CDA) method approach, following [37], to calculate the optical forces from such beams onto a single plasmonic dimer (see Supplementary Information for details on this method).

We select the same angles and refractive indices featured in **Figure 1e** and **Figure 1f** and calculate the longitudinal optical force on a dimer formed by the same rods shown in **Figure 1d** (50 nm wide and deep, 170 and 130 nm long), and explore different radial positions within the beam, enough to cover the higher intensity regions. We also consider two different orientations, one oriented along the beam polarization and one perpendicular to it. Due to the nature of CDA simulations, the dielectric cylinder was omitted, and its refractive index (in both cases) was used as the background index for the CDA simulations.

Results are shown in **Figure 2e** (35º Bessel beam, for the 1.6 refractive index cylinder in **Figure 1e**) and **Figure 2f** (40º Bessel beam, for the 1.7 refractive index cylinder in **Figure 1f**). Both figures display a dual peak shape for the parallel orientation, following closely the shape in **Figure 1c**, with $z$-component forces up to 1 pN. As expected from the single-rod behavior, the magnitude of forces correlates with the positions of higher intensities. On the other hand, the perpendicular orientation, characterized by a lower and blue-shifted polarizability, displays a flatter profile, with forces up to 0.1 pN near 500 nm, and near-zero forces for wavelengths higher than 1000 nm. In particular, the insets of **Figure 2e** and **Figure 2f** show the force profile for radial positions near the edge of their respective cylinders. These positions can be advantageous, enabling the integration of several dimers within the cylinder that can be operated independently. Despite being high-intensity sites, the exerted force is near-zero when oriented orthogonally to polarization, due to their very low and blue-shifted polarizability. Thus, optical pulling is preserved. We highlight that this result showcases the independent physical mechanisms that induce the two types of forces, with no coupling between them. Additional components for the inset cases in **Figure 2e** and **Figure 2f** can be found in the Supplementary Information (**Figure S5**), where it can be seen that, as occurs with a single rod, the transversal forces in dimers can be avoided with a perpendicular orientation with respect to the azimuthal polarization.

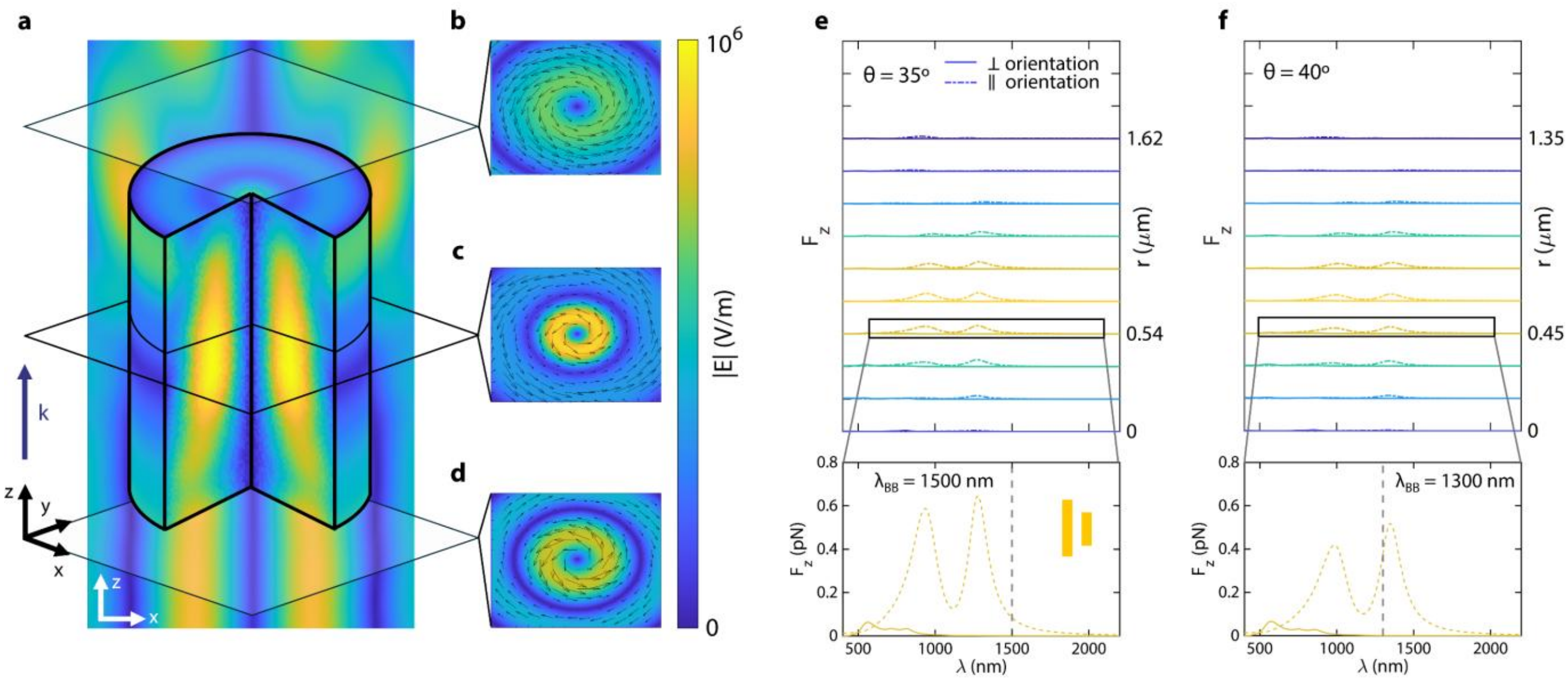

**Figure 2.** a) Electric field within the cylinder in **Figure 1f** (1.44 µm diameter, 1.5 µm), upon interaction with a 40º, 1300 nm Bessel Beam, with cut planes below (b), in the middle (c) and above (d) the cylinder, featuring the field polarization with black arrows. Incidence of the beam is in the positive $z$ direction. e) and f) Wavelength dependence of longitudinal forces exerted on the asymmetric plasmonic dimers (rods 50 nm wide and deep, 170 and 130 nm long, 100 nm gap) by Bessel beams with angles $\theta = 35º$ (e) and $\theta = 40º$ (f). The dimers are located at different radial positions r with respect to the beam center. Solid lines represent an orientation of the rods perpendicular to the azimuthal polarization of the beams while dashed lines represent a parallel orientation. The dimers are placed in a background medium (without the pulling cylinder, hence the positive $F_z$ values) with refractive indices corresponding to those of the cylinders ($n_p = 1.6$ for the 35º beam and $n_p = 1.7$ for the 40º beam).

Due to the axial symmetry of this design, we can position several dimers within the cylinder, making sure that all of them are oriented perpendicular to the Bessel beam polarization. An illustration of this configuration is shown in **Figure 3a**. In this case, the use of 4 dimers allows independent control of movement in the *x* and *y* directions upon linearly-polarized plane wave illumination. This is supported by **Figure 3b-e**, where the optical forces and torques on both cylinders are shown. As expected, the component of force perpendicular to polarization dominates, with a clear peak around 1100 nm. On the other hand, forces on the polarization axis are small, meaning that movement between the two directions can be independently controlled. For both designs, forces are in the 0.1 pN range, sufficient to overcome Brownian motion and obtain a steady movement speed [12].

Looking at the torques, the parallel component dominates, with an oscillating shape around the same wavelength. While the peaks for the perpendicular force component fall within positive torque values (meaning that the cylinder could tumble with time), this effect can be suppressed by using slightly different wavelengths, where the torque can be suppressed.

The longitudinal components of both force and torque ($F_z$ and $T_z$) are provided in the Supplementary Information (**Figure S6**). The longitudinal force is much larger than either $F_{||}$ or $F_{\perp}$, , pointing to strong forward movement when using plane wave illumination. As shown in the Supplementary Information, control of the pushing movement can be done independently of lateral movements if out-of-resonance wavelengths are employed. On the other hand, the longitudinal torque $T_z$ is around 10 times weaker than the $T_{||}$ component in **Figure 3**, meaning that for such illumination $T_{||}$ is the only relevant component.

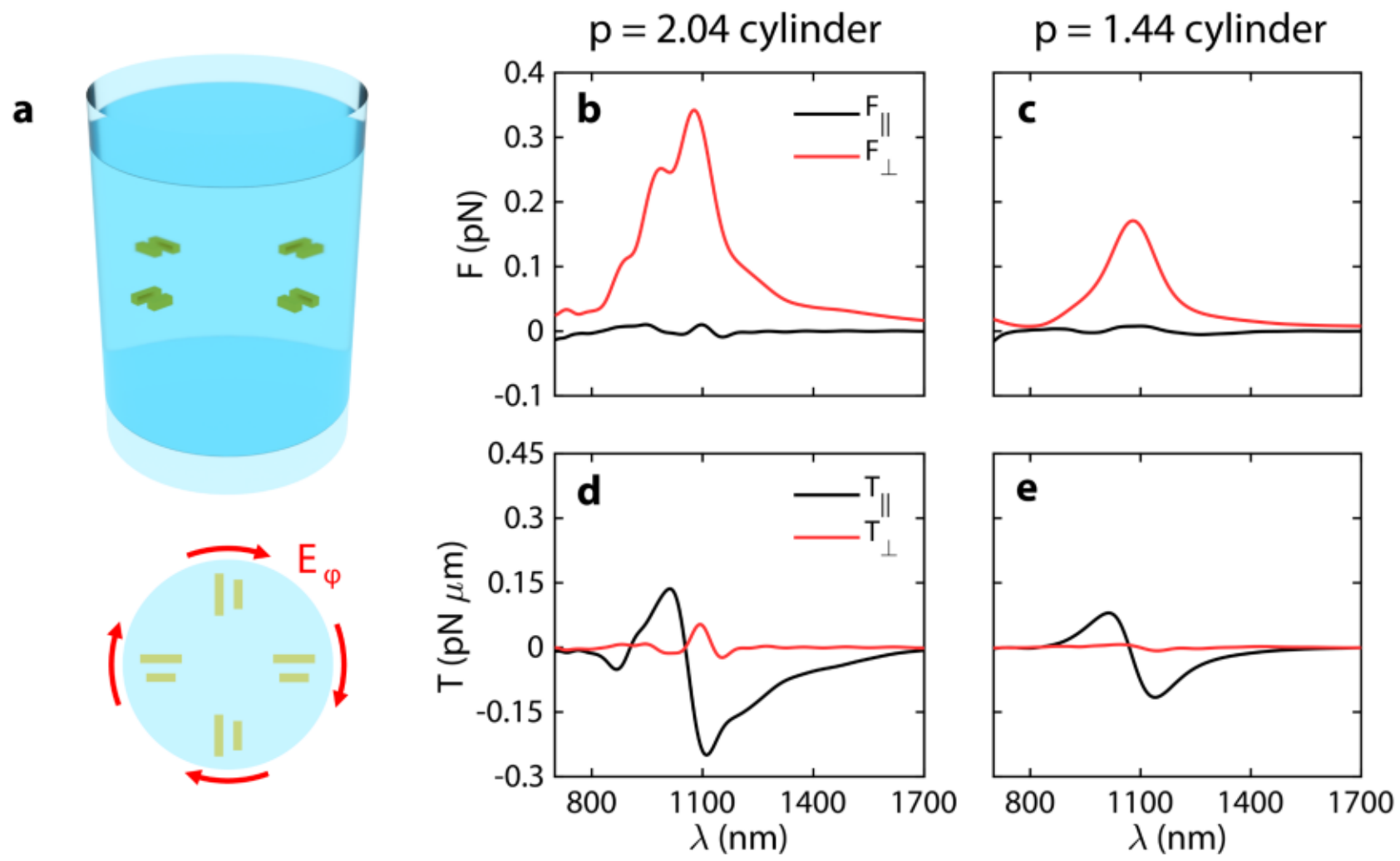

**Figure 3.** a) Scheme of a full 3D nanomotor design, with four plasmonic dimers perpendicularly oriented to the azimuthal polarization of a pulling Bessel beam. b) Optical transversal forces exerted by a 0.4 mW μm$^{-2}$ plane wave on the nanomotors based on cylinders with aspect ratios 2.04 and 1.44 (c). d) Optical transversal torques exerted by a plane wave on the same nanomotors. (e). The cylinder with aspect ratio 2.04 contains three layers of plasmonic dimers, in contrast with the 1.44 aspect ratio cylinder, which only includes one. The background refractive index is $n_b = 1.33$.

The scheme described so far only provides translational motion, but a similar strategy can be employed to implement rotational motion. As shown in the Supplementary Information (**Figure S8**), silver nanorods can be inserted between the dimers. Illumination with circularly polarized light allows transfer of spin angular momentum into the nanorods [38], [39], [40], [41], permitting efficient clockwise and anticlockwise rotations depending on the incident helicity. Furthermore, by placing the rods perpendicular to the azimuthal polarization, the optical pulling force is still preserved.

## 4. Optical pulling stability in diffusion simulations

The longitudinal force maps, shown in **Figure 1e-f**, correspond to cylinders fully aligned with the incident beam. However, when nanomotors are immersed in a fluid such as water, they can undergo Brownian motion, meaning that the pulling must be preserved upon small displacements and rotations from the beam center. Furthermore, as shown in **Figure 3**, interaction with light can induce transversal forces and torques, further prompting the need for stable pulling forces for a realistic application of these nanomotors. In particular, and contrary to optical tweezers, the incident Bessel beams have a ring-shaped maximum intensity, leading to the possibility of gradient optical forces driving the nanomotor away from the beam center [42], [43].

To elucidate the impact of these rotations, we performed FDTD simulations considering displacements from the center along the *x* axis. As the cylinder moves along the *x* axis and away from the center, the incident azimuthal polarization becomes similar to a linear *y*-polarization. Inspired by **Figure 3**, both the *x*-component force and the *y*-component torque will become important and dominate against the other transversal components. Therefore, rotations around the *y* axis are also considered in simulations.

The stability of the optical pulling effect provided by the Bessel beams is showcased in **Figure 4a-b**. While a maximum negative force is found at the beam center, pulling is preserved to rotations up to 20º in both cylinders, and for up to 150-200 nm displacements. The expected dominant transversal components of force and torque are shown in **Figure 4c-d** (transversal force) and **Figure 4e-f** (transversal torque), with insets depicting the non-dominant components. As shown in the insets, the non-dominant components are homogeneous and their contribution can generally be neglected against the dominant components, especially in the case of the longer nanomotor. In contrast, the dominant components of both force and torque resemble those observed in optical tweezers in the ray optics regime [44], leading to equilibrium positions and orientations. The *x*-component force reaches zero around the center of the beam, with the position of this zero slightly shifting to the left or right depending on the nanomotor angular orientation. Similarly, the *y*-component torque reaches zero for most angles at the beam center, and this zero is shifted to angles up to 10º ($p = 2.04$ nanomotor, 40º for the $p = 1.44$ one) outside the beam center. The non-dominant longitudinal torque $T_z$ can be consulted in the Supplementary Information. Once again, the near-zero lateral forces and torques at the beam's positional and angular center highlight the negligible coupling between the optical pulling force and the lateral forces observed in **Figure 3**.

These zero positions and orientations point towards equilibrium points of the nanomotor when illuminated with the Bessel beams. For a further understanding of these restoring forces and torques, we compute potentials from both the dominant transverse forces and torques by integration $U_F(x,\theta) = -\int_{x_{min}}^{x_{max}} F_x(x,\theta)dx$ and $U_T(x,\theta) = -\int_{\theta_{min}}^{\theta_{max}} T_y(x,\theta)\sin|\theta|\,d\theta$. The force-related potential $U_F(x,\theta)$ is represented in **Figure 4g-h**, with the torque-related potential $U_T(x,\theta)$ in the respective insets. The $U_F$ potentials are characterized by two wide potential wells, with centers located around ±180 nm and ±20º (longer design), and ±100 nm and ±30º (shorter design). On the other hand, the $U_T$ potentials follow the zero-line shape of the torque more closely, resulting in potential wells at positive angles for negative *x* positions and negative angles for positive *x* positions. We must highlight that both potentials have values well over 10 $k_BT$ meaning that Brownian motion should have little influence on the overall system dynamics.

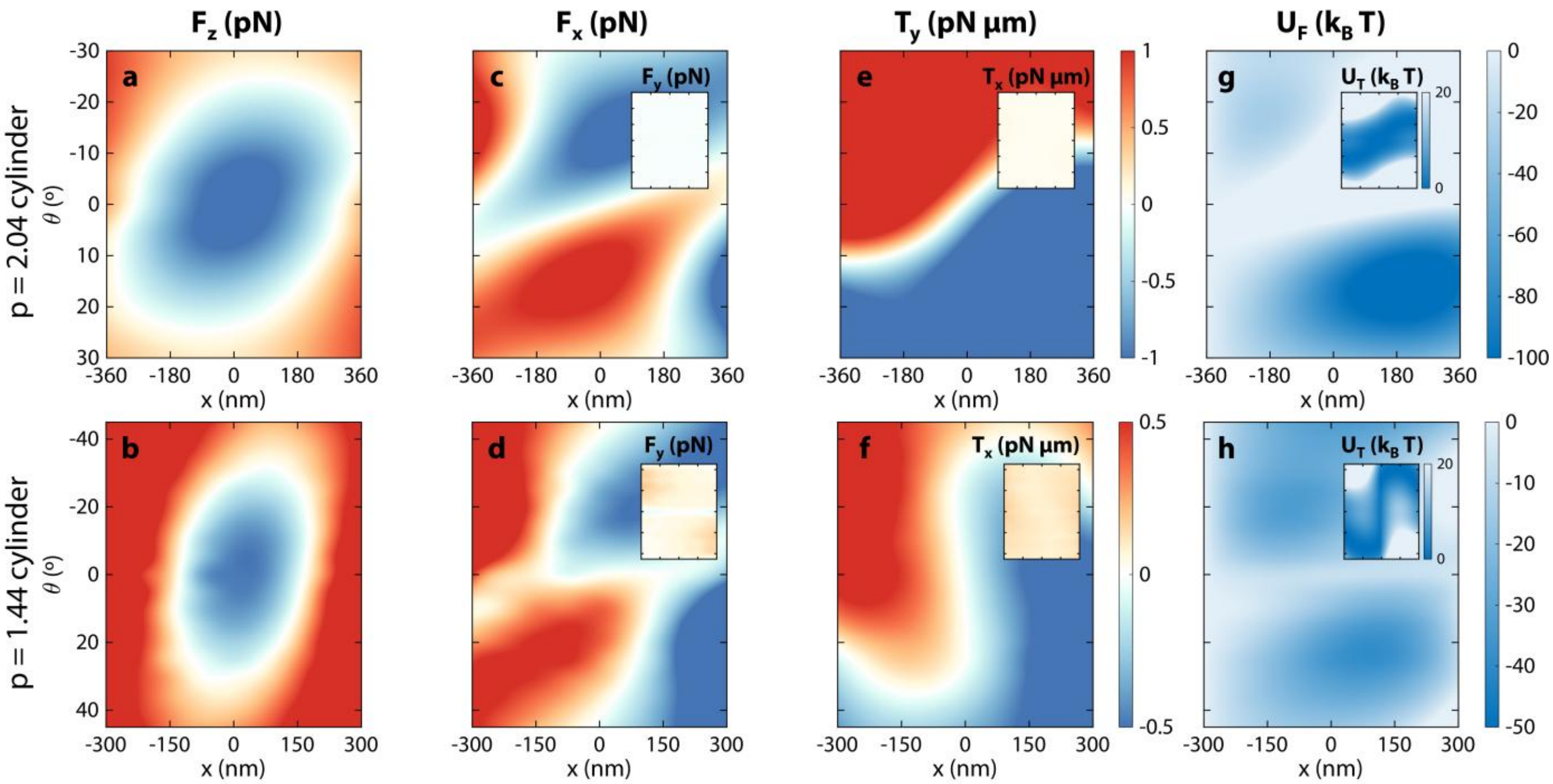

**Figure 4**. a-b) Longitudinal force exerted on the $p = 2.04$ and 1.44 nanomotors by their corresponding Bessel beams at different x positions and tumbling angles $\theta$. c-d) Dominant transversal optical forces (*x*-component), with the negligible *y*-component optical forces shown in the respective insets. e-f) Dominant transversal optical torques (*y*-component), with the negligible *x*-component optical forces shown in the respective insets. g-h) Potential maps associated with the transversal force ($U_F$), with the potential associated to the torques ($U_T$) in the insets. Simulations for this figure include the full 3D-nanomotor system as in **Figure 3** (dielectric cylinder plus plasmonic dimers) in a background medium with refractive index $n_b = 1.33$.

Ultimately, the closeness of the integration domain to the beam center and the difficulty to apply appropriate boundary conditions to the angular integration in $U_T$, disallows adding the two potentials together for a quantitative total potential, meaning that we must make a qualitative interpretation of their shapes.

In principle, the $U_T$ wells, at opposite positions from the ones in $U_F$, should balance a total potential, allowing the nanomotor to travel across all the central region. To address rigorously the non-trivial dynamics in this system, we set up hydrodynamic diffusion simulations (details can be found in the Methods section as well as in the Supplementary Information). We assume an overdamped low-Reynolds-number regime, with external forces given by the dominant components of force and torque from **Figure 4**. To evaluate the stability of the system, we target a long illumination time of 1 second and run 10000 simulations per case in order to obtain significative statistical information.

The overall stability of the nanomotors is demonstrated by the results shown in **Figure 5**. As can be seen in the upper panels of **Figure 5a-b**, nanomotors are distributed around two different positions (one positive, one negative), which is compatible with the two different potential wells in both **Figure 4g** and **Figure 4h**. The longer nanomotor design also follows a similar shape in its angular distribution. However, the shorter design displays a less bi-morph distribution, with a 0º maximum. This can be explained by the similar $U_T$ wells of both designs. As $U_F$ wells are much deeper in the first design, the contribution from $U_T$ does not allow easy passage between the two potential wells.

Meanwhile, the much lower and homogeneous $U_F$ wells for the shorter design point to a joint total potential near the beam center and low angles. This bigger contribution from $U_T$ in the shorter nanomotor also creates a more homogeneous angular profile in the total potential, which is reflected by the wider angular distribution in **Figure 5b**.

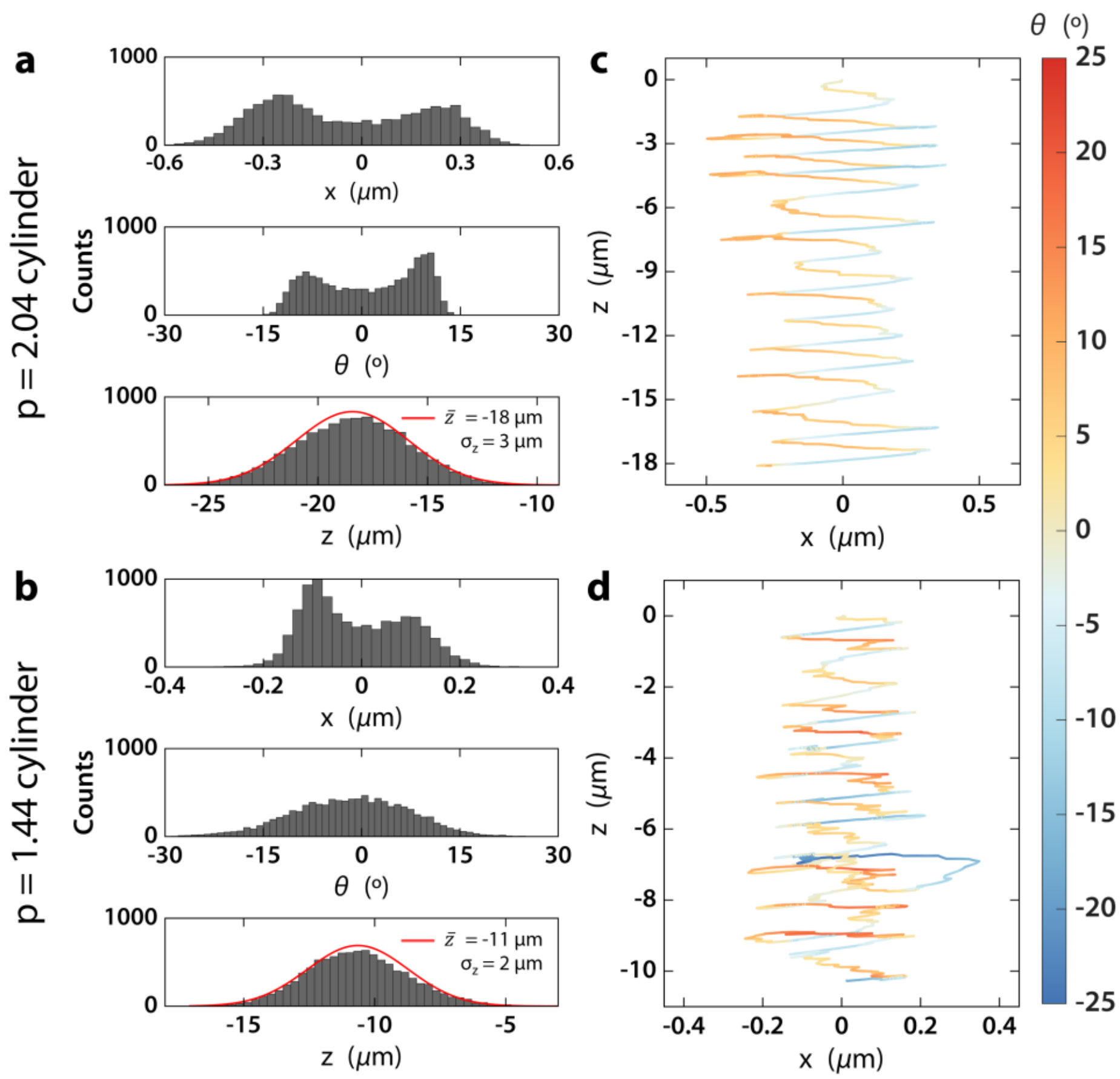

**Figure 5.** Results of hydrodynamic simulations. a-b) Histograms of the final positions, orientations and longitudinal displacement for the $p = 2.04$ (a) and 1.44 (b) aspect ratio nanomotors after 1 s illumination. Gaussian fittings are provided for longitudinal positions. c-d) Typical trajectories of the nanomotors, starting at $(x, \theta, z) = (0,0,0)$ and travelling downwards while oscillating in both $x$ and $\theta$.

This is further confirmed by the typical trajectory of nanomotors, displayed in **Figure 5c-d** and characterized by an oscillation between the two different *x* positions, together with small angular oscillations. This results in the nanomotor being in the pulling region in **Figure 4a-b** for most of the illumination time. This, as shown in the lower panels of **Figure 5a-b**, results in large pulling distances, characterized by Gaussian distributions around 18 and 11 µm, with standard deviations of 3 and 2 µm, respectively. Although the range of both distributions is relatively large (over 10 µm), the standard deviations are comparatively small, leading to moderately narrow distributions. It must be noted that the average values correspond to several times the lengths of the nanomotors, meaning that the speed of nanomotors pulling with moderate intensities is relatively fast. Note that gravitational effects, which could suppose a force contribution close to 0.1 pN (nanomotor masses can be estimated around 10 pg), have been neglected in this analysis. Assuming optical pulling is acting against it, gravitational drag could significantly slow this movement. However, its low magnitude against the optical pulling force means that it can be compensated using a slightly higher illumination intensity.

## 5. Conclusions

In this work, we have revealed how optical pulling can be integrated synergically into transverse light-driven nanomotors, unlocking a new and important degree of freedom to optical manipulation. Optical pulling can be attained employing azimuthally polarized Bessel beam illumination on carefully designed dielectric cylinders, where a FWM is excited and interference at its end produces a collimation of the incident beam. This scheme allows for an efficient generation optical pulling forces, where high-angle Bessel beams are not required and microscale pulling becomes practical. To combine this approach with transverse movements, plasmonic rod dimers were embedded perpendicular to the azimuthal polarization to allow preservation of the optical pulling effect due to the weak plasmonic interaction. Candidate materials for dielectric cylinders that generate optical pulling force include SU-8, Al2O3, and GaN, which are transparent in the visible and near-infrared regions and can be fabricated with high aspect ratios [45], [46], [47]. Additionally, the plasmonic rod dimers can be precisely embedded into the cylinder of these materials through double-exposure electron beam lithography [11], [12], [13], [14]. Experimental realization of Bessel beams with azimuthal polarizations has been done with appropriate metasurfaces [48].

Then, upon illumination with linearly polarized plane waves, the asymmetric scattering from the metallic dimers results in two independent lateral motions, which can be selected with the incident polarization. Pushing motion can also be obtained as a byproduct of the plane wave illumination for lateral movement and can also be independently controlled by shifting the incident wavelength. As demonstrated in the Supplementary Information, complementary structures can be implemented into the design to efficiently provide controlled rotations [33], [38], [39], [40], [41], which can be independently controlled from other degrees of freedom upon tuning their plasmonic resonances to a different wavelength than the one for translation movement. This can be accomplished using different sizes or materials [36].

Finally, we have demonstrated how this approach is robust enough against rotations, displacements and misalignments, which can be caused by either Brownian motion, the complex Bessel beam profile, or a combination of both. The restoring nature of the transversal forces and torques exerted by the beam produces an intricate potential profile, where the nanomotors oscillate around the beam center while maintaining a straight orientation over long times. This stability, together with the low intensities required for manipulation, greatly enhances the applicability of our approach. We note that other nanomotor designs are characterized by lower height/lateral size aspect ratios (for example, $p < 0.1$ in [13]). However, these aspect ratios do not consider the strict conditions imposed by optical pulling forces. Nonetheless, this could also be reached by our design, if the wavelength of the Bessel beam is drastically lowered. For that, the plasmonic dimers should be tuned to UV wavelengths, for which the use of materials different than gold would be required [49], [50].

## 6. Methods

### Numerical electrodynamical simulations

All numerical electromagnetic calculations in this work, excepting those specified with the Couple Dipole Approximation approach in **Figure 2** (a detailed discussion can be found in the Supplementary Material), were performed by 3D full-wave electrodynamic simulation software Lumerical FDTD, providing an accurate solution of the Maxwell equations. The non-diffracting Bessel beams were implemented into FDTD using a combination of 60 different Total Field/Scattered Field (TFSF) sources, following [41] (more details in the Supplementary Material). The angular dispersion for broadband sources in Lumerical FDTD meant that simulations had to be single-frequency to provide accurate results. This motivated us to use the CDA approach for broadband simulations when considering the plasmonic dimer. Calculation of optical forces $\boldsymbol{F}$ and torques $\boldsymbol{T}$ was done following the Maxwell stress tensor formalism [42]:

$$\boldsymbol{F} = \oiint_{\Sigma} \overline{\boldsymbol{T_M}} \cdot \boldsymbol{n} \, d\Sigma \tag{2}$$

$$\boldsymbol{T} = -\oiint_{\Sigma} (\overline{\boldsymbol{T_M}} \times \boldsymbol{r}) \cdot \boldsymbol{n} \, d\Sigma \tag{3}$$

where $\boldsymbol{r}$ is the position vector, $\boldsymbol{n}$ is the outward normal unit vector and the integration is performed over a surface $\Sigma$ enclosing the nanomotor. $\overline{\boldsymbol{T_M}}$ is the time-averaged Maxwell stress tensor:

$$\overline{\boldsymbol{T_M}} = \frac{1}{2}\Re\mathrm{e}\left[\ \varepsilon_r\varepsilon_0\boldsymbol{E}\otimes\boldsymbol{E}^* + \mu_0\boldsymbol{H}\otimes\boldsymbol{H}^* - \frac{1}{2}(\varepsilon_r\varepsilon_0\boldsymbol{E}\otimes\boldsymbol{E}^* + \mu_0\boldsymbol{H}\otimes\boldsymbol{H}^*)\,\overline{\boldsymbol{I}}\ \right] \tag{4}$$

where $\varepsilon_r$ is the relative permittivity, $\otimes$ denotes the dyadic product, $*$ the complex conjugate and $\overline{\boldsymbol{I}}$ is the identity dyadic.

## Hydrodynamic simulations

Hydrodynamic simulations were done following the formalism for diffusion of non-spherical particles in [42]. Due to the axial symmetry and the prevalence of $x$-component forces and $y$-component torques against other components, diffusion along the $x$ and $z$ axes, and increments in the $\beta$ angle (tumbling along the $x$ axis) were considered. Assuming a low-Reynolds-number regime, with external forces $F_x$, $F_z$ and torques $T_y$, the movement of the particle after a time step $\Delta t$ will be given by the overdamped Langevin equation

$$\begin{bmatrix}\Delta x_p\\ \Delta z_p\\ \Delta\beta_p\end{bmatrix} = \frac{\overline{\boldsymbol{D}}}{k_B T}\Delta t\begin{bmatrix}F_{x,p}\\ F_{z,p}\\ T_{y,p}\end{bmatrix} + \sqrt{2\Delta t}\begin{bmatrix}w_x\\ w_z\\ w_\beta\end{bmatrix} \tag{5}$$

where the $p$ subindex denotes the particle's frame of reference. $k_B$ is the Boltzmann constant and $T$ is the environment temperature. $\overline{\boldsymbol{D}}$ is the particle's characteristic diffusion tensor, which for cylindrical particles in the considered dimensions has the form

$$\overline{\boldsymbol{D}} = \begin{bmatrix}D_t & 0 & 0\\ 0 & D_t & 0\\ 0 & 0 & D_r^\perp\end{bmatrix} \tag{6}$$

Following [43], each of the components of this diffusion tensor will be strongly dependent on the cylinder aspect ratio $p$, and can be extracted using interpolating equations. Finally, $w_x$, $w_z$ and $w_\beta$, accounting for Brownian motion; correspond to a set of random numbers extracted from a multivariate normal distribution with mean zero and covariance $\overline{\boldsymbol{D}}$.

For each time step ($\Delta t = 0.001$ s), external forces and torques are extracted from FDTD results based on the particle's position and orientation, and converted to the particle's frame of reference. Particle position and orientation are updated based on equation 5, and then the particle's frame of reference is updated accordingly.

We note that the overdamped approximation in equation 5 must be justified by the ratio of the characteristic time of an optical trap $\tau_{OT} = \gamma/\kappa$ and the momentum relaxation time $\tau_m = m_{NM}/\gamma$, where $\gamma = \frac{k_B T}{D_t}$ is the friction coefficient, $\kappa$ is the trap stiffness and $m_{NM}$ is the nanomotor mass [42]. As can be derived from the Supplementary Material, $\gamma \sim 10^{-8}$ kg/s; while the nanomotor masses can be estimated to $m_{NM} \sim 10^{-14}$ kg by approximating them to glass cylinders with density $\rho = 2500$ kg/m$^3$. Estimation of the trap stiffness is not trivial, given the complex dependence of forces and torques on position and angle. However, the force magnitude in the pN range, compared to $10^{-7}$ m displacements, suggests stiffnesses of up to around $10^{-5}$ N/m. These estimated quantities yield approximately characteristic times $\tau_{OT} \sim 10^{-3}$ s, much higher than the momentum relaxation time $\tau_m \sim 10^{-6}$ s, thus justifying the overdamped treatment and the choice of the time step.

**Author contributions.** All authors contributed to the conceptualization of the study. G. S. carried out numerical simulations, formal analysis, and writing of the original draft. Y. Y. T. and P. A. participated in the discussion of results and supervised the project. All authors participated in reviewing and editing the manuscript.

**Funding.** This work acknowledges funding by the MOPHOSYS Project (PID2022-139560NB-I00) from Proyectos de Generación de Conocimiento provided by the Spanish Agencia Estatal de Investigación. This work was supported by Grants-in-Aid for Scientific Research (KAKENHI) (Nos. JP24H00424 and JP22H05132 in Transformative Research Areas (A) "Chiral materials science pioneered by the helicity of light" to Y.Y.T.) from the Japan Society for the Promotion of Science (JSPS), and JST FOREST Program (No. JPMJFR213O to Y.Y.T.).

**Acknowledgments.** G. S. thanks the Spanish Ministry of Education for his predoctoral contract grant (FPU21/02296). G. S. would also like to thank Y. Gutiérrez for her assistance in the development of the CDA calculation method.

**Disclosure.** The authors declare no conflict of interest.

**Supplementary material.** Supplementary Material contains additional details on: excitation of the FWM and additional suitable designs, antireflection coatings design, implementation of Bessel beams in FDTD, CDA method for optical force calculations in plasmonic rod dimers, longitudinal component forces and torques on nanomotors, and characterization of the hydrodynamic properties of cylinders.

# Supplementary Information

## 1. Excitation of the FWM and additional suitable designs

We follow the basic theory of waveguide coupling [1] and apply it to the dielectric cylinder as done in [2]. For such a cylinder, both TM and TE waveguide modes. Particularly, an $l-$order TE mode in a waveguide with radius a will be described by a tangential electric field

$$E_{\varphi}(\rho,\varphi,z) = \left[-\frac{ik_z a^2 l}{u\rho^2} A J_l\left(\frac{u\rho}{a}\right) + \frac{i\omega\mu a}{u} B J'_l\left(\frac{u\rho}{a}\right)\right]\cos(l\varphi)\, e^{ik_z z} \qquad (S1)$$

where $k_z$ is the propagation constant, $\omega$ is the angular frequency of the wave and $u^2 = \left(k_p - k_z^2\right)a^2$and $A$ and $B$ are constants. For $l = 0$, this transforms into

$$E_{\varphi}(\rho,\varphi,z) = \frac{i\omega\mu a}{u} B J'_0\left(\frac{u\rho}{a}\right) e^{ik_z z} \qquad (S2)$$

which has a similar spatial profile as the azimuthally polarized Bessel beam (itself a TE wave, characterized by a purely transversal $E_{\varphi}$) in equation 1 in the main text. This similar polarization and spatial profile mean that the coupling efficiency $\eta$, expressed through the overlap integral at the waveguide cross-sectional surface $\Sigma$ [3]

$$\eta \propto \left|\iint_{\Sigma} \boldsymbol{E}^{*}_{\boldsymbol{BB}}(\rho,\varphi)\cdot \boldsymbol{E}_{\boldsymbol{TE}}(\rho,\varphi)\, d\Sigma\right|^2 \qquad (S3)$$

will be maximized. This maximum coupling, in turn, suppresses photon recoil, favoring the optical pulling effect. A similar phenomenon occurs when coupling a radially polarized (TM) Bessel beam with a TM-supporting waveguide, hence the viability of radially polarized beams for optical pulling with small-angle Bessel beams.

Thus, cylinders must be able to support the TE mode in order to maximize the pulling effect. The cutoff condition for such excitation is given by the solution of the equation

$$\frac{J_1(u)}{J_0(u)} = -\frac{u}{w}\frac{K_1(w)}{K_0(w)} \qquad (S4)$$

where $w = \left(k_z^2 - k_b\right)a^2$ and $K_l(x)$ is the $l-$order modified Bessel function of the second kind. For an incident Bessel beam with incidence angle $\theta_0$ , $k_z = k_p\cos(\theta_p)$, and for small $\theta_p$, an approximate solution is given by

$$D(\theta_0) = \frac{2z_1}{k_b\theta_0} \qquad (S5)$$

where $z_1$ is the first zero of the Bessel function $J_1(u)$. However, greater incidence angles require a full numerical solution of equation S4. This is evidenced in **Figure S1**, where the solution of equation S4 for a wavelength of 1500 nm and indices $n_b = 1.33$ and $n_p = 1.6$ is shown, together with the approximation in equation S5. While the overall shapes of the solutions are similar, and values for small angles have little discrepancy, this difference becomes important once high angles are reached. The inset zooms in both solutions for angles between 30 and 40º, where the difference between solutions is approximately 0.5 µm, around 30% in relative error.

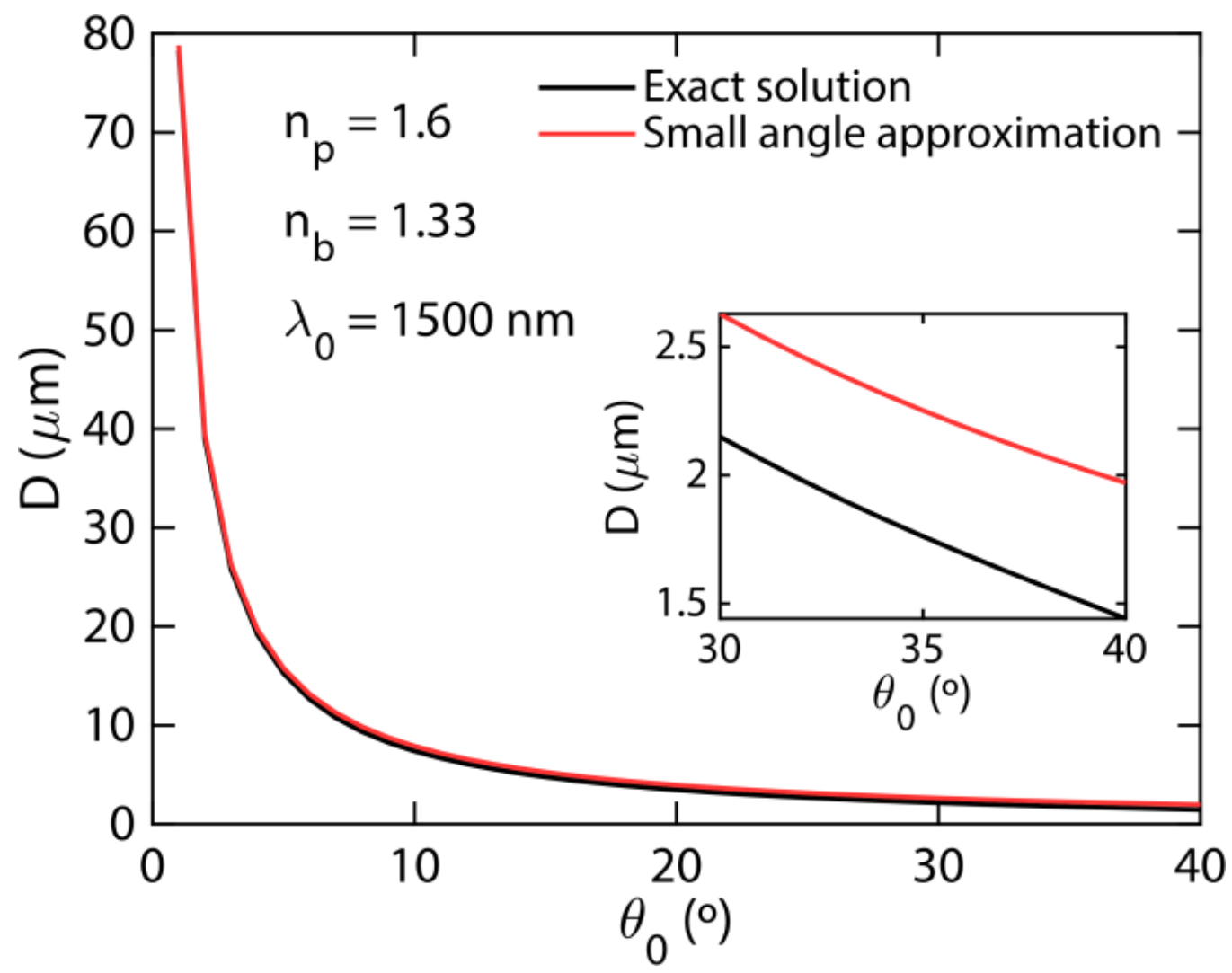

**Figure S1.** Comparison of the solution of equation S4 and equation S5 for a wavelength of 1500 nm and indices combination $n_b = 1.33$ and $n_p = 1.6$. The inset zooms in the $\theta_0 \in [30{,}40]^\circ$ range, where differences can be appreciated better. The 1.77 µm diameter used for the cylinder in **Figure 1e** in the main text can be extracted from the black line at $\theta_0 = 35^\circ$.

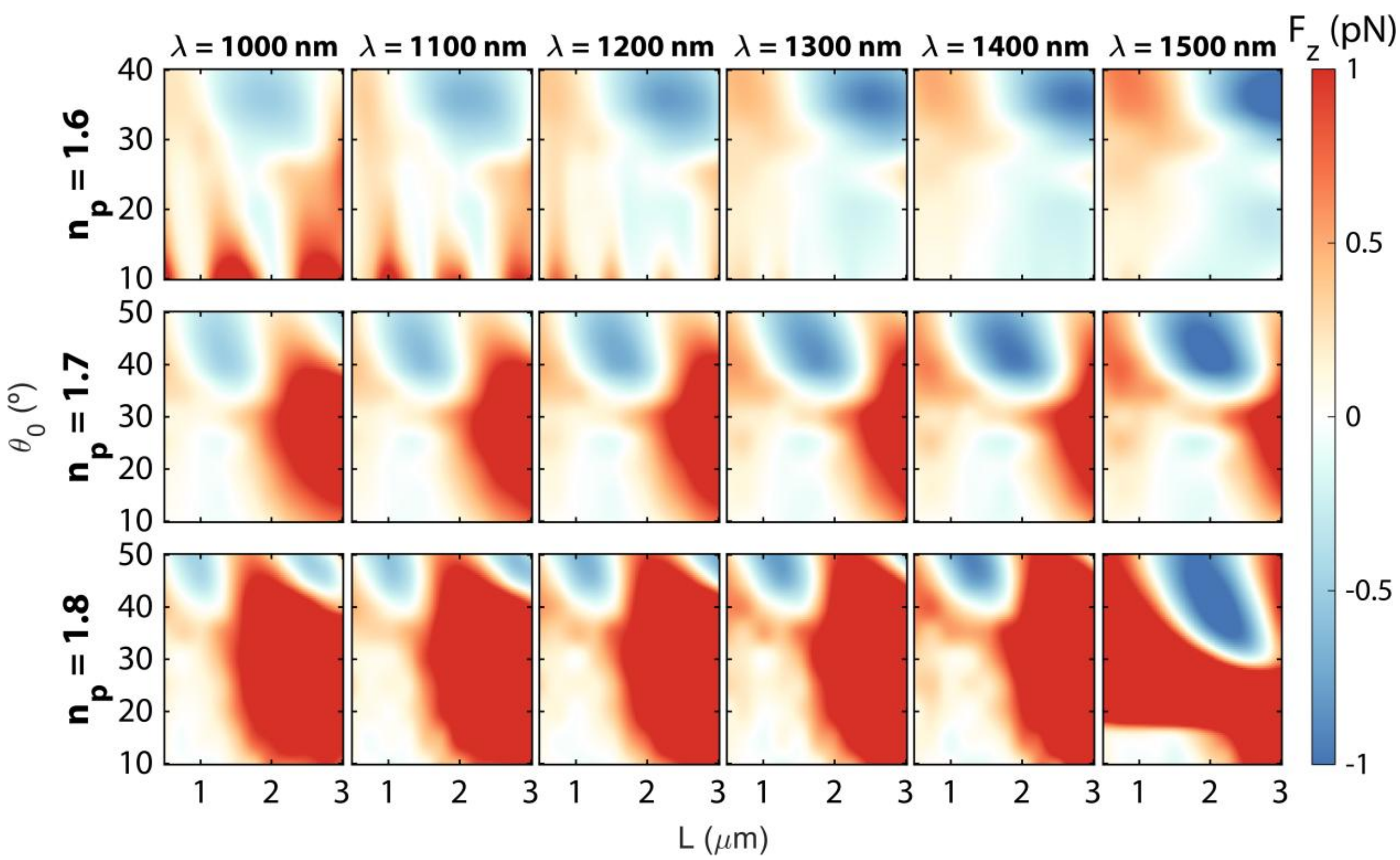

**Figure S2.** Phase maps of longitudinal optical force for the whole explored casuistry. The background index is $n_b = 1.33$ for all cases.

Not all beams can be coupled to the mode. For each refractive index combination, there exists a critical angle $\theta_{crit} = \mathrm{asin}\left[\sqrt{\left(\frac{n_p}{n_b}\right)^2 - 1}\right]$ over which the FWM cannot be excited, putting an upper bound to the exploration of different cone angles [4]. For example, for $n_p = 1.6$ and $n_b = 1.33$, $\theta_{crit} \approx 42^\circ$, and for $n_p = 1.5$; it drops to $\theta_{crit} \approx$ 31º. On the other hand, angles over 45º quickly suppress the range of the Bessel beam. Therefore, our search has limited to $n_p \in [1.6, 1.8]$ and angles $\theta_0 \in [10, 50]$º, although angles over 40º have been discarded in the $n_p = 1.6$ case due to being close to the critical angle.

**Figure S2** shows the phase maps for the whole explored casuistry. In general, longer wavelengths and greater refractive index correspond to stronger pulling forces at their minima. High refractive index contrasts also help to bring the minima to shorter disks. For the shorter wavelengths and higher refractive indices the oscillation of the force at high angles, between pulling and pushing, becomes more visible. This is due to interference, and the period of the oscillation follows $\Delta = \frac{\lambda}{n_p - n_b}$ [5]. Also, pulling force at lower angles becomes more viable at lesser refractive index contrasts, as reflections are suppressed more efficiently by the ARCs.

## 2. Antireflection Coatings design

To improve the pulling efficiency of our nanomotors, each of the cylinders analyzed in this work have been equipped with antireflection coatings (ARCs) at both ends. This strategy has been successfully employed in other works [6], [7]. The refractive index of an ARC is given by

$$n_{ARC} = \sqrt{n_p n_b} \qquad (S6)$$

while its thickness is given by

$$d(\theta_0) = \frac{\lambda \cos \theta_{ARC}}{4(n_{ARC} - n_b \sin \theta_0 \sin \theta_{ARC})} \qquad (S7)$$

where the propagation angle within the ARC, $\theta_{ARC}$, is given by Snell's law:

$$\theta_{ARC} = \mathrm{asin}\left(\frac{n_b \sin \theta_0}{n_{ARC}}\right) \qquad (S8)$$

Thus, each of the refractive combinations and wavelengths (each panel in **Figure S2**), requires ARCs with different refractive indices and/or thicknesses.

## 3. Implementation of Bessel beams in FDTD

Due to the high cone angles considered in this work, a non-paraxial implementation of Bessel beams in FDTD was needed. The approach described in [8] was followed. Any arbitrary Bessel beam can be understood, in the Angular Spectrum Representation (ASR), as a superposition of plane waves with wavevectors lying on a conical surface (defined by the cone angle $\theta_0$). In a spherical coordinate system with position vector $\boldsymbol{r}$, at the focal point of a lens, the field of an $m-$order Bessel beam propagating along z is

$$\boldsymbol{E}(\boldsymbol{r}) = \frac{ikfe^{-ikf}}{2\pi} \int_{\theta}^{\theta_{max}} \int_{0}^{2\pi} \boldsymbol{E_{PW}}(\theta, \varphi) e^{i\boldsymbol{k}\cdot\boldsymbol{r}} \sin\theta \, d\theta d\varphi \qquad (S9)$$

where $f$ is the focal length of the lens, and the wavevector $\boldsymbol{k} = (k \sin\theta \cos\varphi, k \sin\theta \sin\varphi, k \cos\theta)$. The angular spectrum function $\boldsymbol{E_{PW}}$ is given by [9]

$$\boldsymbol{E_{PW}}(\theta, \varphi) = E_{PW0}(\theta_0, \varphi) e^{im\varphi} \frac{\delta(\theta - \theta_0)}{\sin\theta} \boldsymbol{Q}(\theta_0, \varphi) \qquad (S10)$$

where $E_{PW0}$ is the amplitude of each plane wave that contributes to the beam, $\delta(\theta - \theta_0)$ is the Dirac delta distribution, limiting contributions to the cone surface with angle $\theta_0$; and the $\boldsymbol{Q}(\theta_0, \varphi)$ vector describes polarization as [10]

$$\boldsymbol{Q}(\theta_0,\varphi) = \begin{bmatrix} p_X(\cos\theta_0\cos^2\varphi + \sin^2\varphi) - p_y(1-\cos\theta_0)\sin\varphi\cos\varphi \\ -p_X(1-\cos\theta_0)\sin\varphi\cos\varphi + p_y(\cos\theta_0\sin^2\varphi + \cos^2\varphi) \\ -p_x\sin\theta_0\cos\varphi - p_y\sin\theta_0\sin\varphi \end{bmatrix} \quad (S11)$$

and depends on the pair of parameters $(p_x, p_y)$. An azimuthal polarization is given by $(p_x, p_y) = (-\sin\varphi, \cos\varphi)$. Then, the implementation of Bessel beams consists in introducing a combination of plane waves whose amplitude, direction and polarization matches the angular spectrum function $\boldsymbol{E_{PW}}$. The finiteness of the number of plane waves means that the integral in equation S9 will be discretized as

$$\boldsymbol{E}(\boldsymbol{r}) = \sum_{n=0}^{N} C_n E_{PW0}\, \boldsymbol{Q}(\theta_0, \varphi_n) e^{im\varphi} e^{i\boldsymbol{k}\cdot\boldsymbol{r}} \quad (S12)$$

where $\varphi_n$ corresponds to the azimuthal angle integration points, weighted by coefficients $C_n$. They can be determined as

$$\varphi_n = \frac{2\pi n}{N}, \qquad n = 0,1 \dots N$$
$$C_n = \begin{cases} \frac{\pi}{N}, & n = 0, N \\ \frac{2\pi}{N}, & n = 1,2 \dots N-1 \end{cases} \quad (S13)$$

where the half-valued $C_n$ coefficients at $n = 0, N$ compensate the degeneracy at that point. To introduce these plane waves without periodic boundary conditions in FDTD, the Total Field/Scattered Field (TFSF) within the Lumerical FDTD software approach is followed.

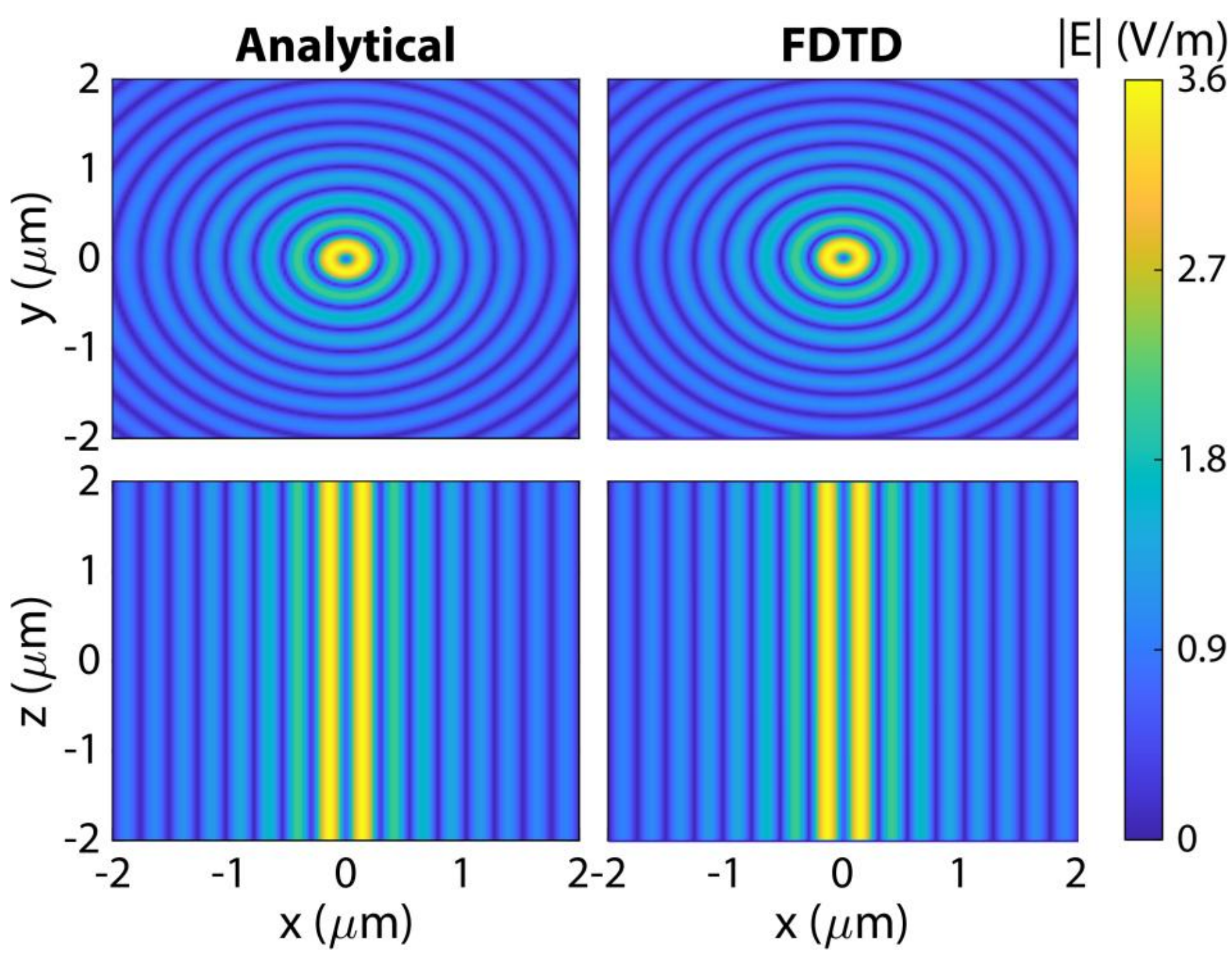

**Figure S3.** Comparison of analytical and FDTD-generated Bessel beams, at $XY$ and $XZ$ planes. The beam is zero-order with azimuthal polarization, and is characterized by an angle of 55º, a wavelength of 532 nm, and the plane wave amplitude is set to $E_{PW0} = 1$ V/m. The surrounding medium has an index $n_b = 1.33$.

As shown in [8], a combination of 60 plane wave sources is enough to closely match the analytical profile of the beam. Typically, injection angles in broadband simulations change as a function of frequency, with only the center frequency matching the nominal injection angle. While this dispersion can be negligible for small angles, steep incidences cause notable differences. This effect can be corrected using the Broadband Fixed Angle Source Technique (BFAST) or using frequency dependent profiles [11], these approaches are not implemented for TSFS sources, meaning that simulations had to be limited to a single frequency, matching the nominal angle. **Figure S3** displays the excellent match between the analytical Bessel beam from equation 1 and the TFSF approach in Lumerical FDTD for a zero-order azimuthally polarized Bessel beam with $\theta_0 = 55^{\circ}$ at a wavelength of 532 nm.

## 4. Derivation of the force exerted by a zero-order azimuthally polarized Bessel beam on a plasmonic particle

We start by considering the Lorentz force on a point dipole $\boldsymbol{p}$, given by [12], [13]:

$$\langle \boldsymbol{F} \rangle = \frac{1}{2} \Re\mathrm{e}[(\boldsymbol{p}^* \cdot \nabla)\boldsymbol{E} + i\omega(\boldsymbol{p}^* \times \boldsymbol{B})] \tag{S14}$$

where $\boldsymbol{E}(\rho, \varphi, z) = -iE_0 e^{ik_0 \cos\theta_0 z} \frac{J_0'(k_0 \sin\theta_0 \rho)}{\sin\theta_0} \boldsymbol{n_\varphi} = E_\varphi \boldsymbol{n_\varphi}$, $\boldsymbol{p} = \varepsilon_\mathrm{b} \alpha \boldsymbol{E}$ (where $\varepsilon_\mathrm{b} = \varepsilon_0 n_b^2$) and $\boldsymbol{B} = \frac{1}{i\omega} \nabla \times \boldsymbol{E}$. We can substitute and obtain:

$$\langle \boldsymbol{F} \rangle = \frac{1}{2} \Re\mathrm{e}[\varepsilon_\mathrm{b} \alpha^* (\boldsymbol{E}^* \cdot \nabla)\boldsymbol{E} + \alpha^*(\boldsymbol{E}^* \times (\nabla \times \boldsymbol{E}))] \tag{S15}$$

Because the field only has the azimuthal component, the dot product in the first term simplifies to

$$(\boldsymbol{E}^* \cdot \nabla)\boldsymbol{E} = E_\varphi^* \frac{1}{\rho} \frac{\partial E_\varphi \boldsymbol{n_\varphi}}{\partial \varphi} = E_\varphi^* \frac{1}{\rho} \left( \frac{\partial E_\varphi}{\partial \varphi} \boldsymbol{n_\varphi} - E_\varphi \boldsymbol{n_\rho} \right) = -\frac{|E_\varphi|^2}{\rho} \boldsymbol{n_\rho} \tag{S16}$$

where in the last step the fact that the field has cylindrical symmetry $\left( \frac{\partial E_\varphi}{\partial \varphi} = 0 \right)$ has been exploited. For the other term in the equation, we can first compute $\nabla \times \boldsymbol{E}$:

$$\nabla \times \boldsymbol{E} = \left( -\frac{1}{\rho} \frac{\partial(\rho E_\varphi)}{\partial z}, 0, \frac{1}{\rho} \frac{\partial(\rho E_\varphi)}{\partial \rho} \right) \tag{S17}$$

With this, the product $\boldsymbol{E}^* \times (\nabla \times \boldsymbol{E})$ yields:

$$\boldsymbol{E}^* \times (\nabla \times \boldsymbol{E}) = \begin{vmatrix} \boldsymbol{n_\rho} & \boldsymbol{n_\varphi} & \boldsymbol{n_z} \\ 0 & E_\varphi^* & 0 \\ -\frac{1}{\rho} \frac{\partial(\rho E_\varphi)}{\partial z} & 0 & \frac{1}{\rho} \frac{\partial(\rho E_\varphi)}{\partial \rho} \end{vmatrix} = E_\varphi^* \frac{1}{\rho} \frac{\partial(\rho E_\varphi)}{\partial \rho} \boldsymbol{n_\rho} + E_\varphi^* \frac{1}{\rho} \frac{\partial(\rho E_\varphi)}{\partial z} \boldsymbol{n_z} \tag{S18}$$

The first derivative can be simplified as $\frac{\partial(\rho E_\varphi)}{\partial \rho} = E_\varphi + \rho \frac{\partial E_\varphi}{\partial \rho}$, and the total force becomes

$$\langle \boldsymbol{F} \rangle = \frac{1}{2} \Re\mathrm{e}\left[ -\varepsilon_\mathrm{b} \alpha^* \frac{|E_\varphi|^2}{\rho} \boldsymbol{n_\rho} + \varepsilon_\mathrm{b} \alpha^* E_\varphi^* \frac{1}{\rho} \left( E_\varphi + \rho \frac{\partial E_\varphi}{\partial \rho} \right) \boldsymbol{n_\rho} + \varepsilon_\mathrm{b} \alpha^* E_\varphi^* \frac{1}{\rho} \frac{\partial(\rho E_\varphi)}{\partial z} \boldsymbol{n_z} \right] \tag{S19}$$

which simplifies to

$$\langle \boldsymbol{F} \rangle = \frac{1}{2}\Re\mathrm{e}\left[-\varepsilon_\mathrm{b}\alpha^* \frac{|E_\varphi|^2}{\rho}\boldsymbol{n_\rho} + \varepsilon_\mathrm{b}\alpha^* \frac{|E_\varphi|^2}{\rho}\boldsymbol{n_\rho} + \varepsilon_\mathrm{b}\alpha^* \frac{\partial E_\varphi}{\partial \rho}\boldsymbol{n_\rho} + \varepsilon_\mathrm{b}\alpha^* E_\varphi^* \frac{1}{\rho}\frac{\partial(\rho E_\varphi)}{\partial z}\boldsymbol{n_z}\right] =$$
$$= \frac{1}{2}\Re\mathrm{e}\left[\varepsilon_\mathrm{b}\alpha^* E_\varphi^* \left(\frac{\partial E_\varphi}{\partial \rho}\boldsymbol{n_\rho} + \frac{1}{\rho}\frac{\partial(\rho E_\varphi)}{\partial z}\right)\boldsymbol{n_z}\right] \quad \text{(S20)}$$

which is equation 3 in the main text. Further simplification can be achieved by computing the necessary derivatives:

$$\frac{\partial(\rho E_\varphi)}{\partial z} = -\rho E_0 e^{ik_0 \cos\theta_0 z} \ \frac{J_1(k_0 \sin\theta_0\, \rho)}{\sin\theta_0} \ k_0 \cos\theta_0 \quad (S21a)$$

$$\frac{\partial E_\varphi}{\partial \rho} = iE_0 e^{ik_0 \cos\theta_0 z} \ \frac{J_1'(k_0 \sin\theta_0\, \rho)}{\sin\theta_0} \ k_0 \sin\theta_0 = ik_0 E_0 e^{ik_0 \cos\theta_0 z} J_1'(k_0 \sin\theta_0\, \rho) \quad (S21b)$$

where we have exploited the recurrence relation of Bessel's functions: $J_m'(x) = \frac{1}{2}[J_{m-1}(x) - J_{m+1}(x)]$. For $m = 0$, $J_0'(x) = \frac{1}{2}[J_{-1}(x) - J_1(x)] = \frac{1}{2}[-J_1(x) - J_1(x)] = -J_1(x)$. Inserting these expressions together with $E_\varphi$ yields the final expression:

$$\langle \boldsymbol{F} \rangle = \frac{1}{2}\Re\mathrm{e}\left[\varepsilon_\mathrm{b}\alpha^* E_0^2 \left(\frac{k_0}{\sin\theta_0} J_1(x_\rho) J_1'(x_\rho)\boldsymbol{n_\rho} + i\frac{k_0\cos\theta_0}{\sin^2\theta_0} J_1^2(x_\rho)\boldsymbol{n_z}\right)\right] \quad (S22)$$

which is equation 4 in the main text, with $x_\rho = k_0 \sin\theta_0\, \rho$. As stated there, the negative sign of the imaginary part of the conjugate polarizability $\alpha^*$ will yield a positive longitudinal force. Interestingly, the $\frac{\cos\theta_0}{\sin^2\theta_0}$ means that this longitudinal force will be greater for small-angle Bessel beams, which resemble plane waves. On the other hand, when $\theta_0 \to \frac{\pi}{2}$, the longitudinal force approaches zero.

## 5. CDA method for optical force calculations in plasmonic rod dimers

We follow the method outlined in [14], although we note the use of SI units below. Each of the rods in the dimer is modeled as a point dipole $\boldsymbol{p_i}$ characterized by a polarizability $\alpha_i$. Under this approximation, the optical response of a dimer in a medium with dielectric constant $\varepsilon_b = \varepsilon_0 n_b^2$ can be obtained by solving a system of equations:

$$\begin{aligned} \boldsymbol{p_1} &= \varepsilon_b\alpha_1\boldsymbol{E_1} = \varepsilon_b\alpha_1(\boldsymbol{E_{inc,1}} - \overline{\boldsymbol{A_{12}}}\cdot\boldsymbol{p_2}) \\ \boldsymbol{p_2} &= \varepsilon_b\alpha_2\boldsymbol{E_2} = \varepsilon_b\alpha_2(\boldsymbol{E_{inc,2}} - \overline{\boldsymbol{A_{21}}}\cdot\boldsymbol{p_1}) \end{aligned} \quad (S23)$$

where $\boldsymbol{E_i}$ corresponds to the total electric field at the position of dipole $\boldsymbol{p_i}$, $\boldsymbol{E_{inc,1}}$ is the incident field on that position and the product $\overline{\boldsymbol{A_{ij}}}\cdot\boldsymbol{p_j}$ gives the electric field caused by dipole $\boldsymbol{p_j}$ at the position of dipole $\boldsymbol{p_i}$. $\overline{\boldsymbol{A_{ij}}}$ is a second order tensor given by

$$\overline{\boldsymbol{A_{ij}}}(\boldsymbol{r_{ij}}) = \frac{e^{ikr_{ij}}}{4\pi\varepsilon_b r_{ij}}\left[k^2(\boldsymbol{n_{ij}}\otimes\boldsymbol{n_{ij}} - \overline{\boldsymbol{I}}) + \frac{ikr_{ij}-1}{r_{ij}^2}(3\boldsymbol{n_{ij}}\otimes\boldsymbol{n_{ij}} - \overline{\boldsymbol{I}})\right] \quad (S24)$$

where $\boldsymbol{r_{ij}}$ is the position vector separating the dipoles. $r_{ij} = |\boldsymbol{r_{ij}}|$, $\boldsymbol{n_{ij}} = \frac{\boldsymbol{r_{ij}}}{r_{ij}}$ and $\overline{\boldsymbol{I}}$ is the identity dyadic. The symbol $\otimes$ denotes the dyadic product. For two dipoles, the system S11 can be easily solved as

$$\boldsymbol{p_1} = \frac{\varepsilon_b \alpha_1 \left( \boldsymbol{E_{inc,1}} - \varepsilon_b \alpha_2 \, \overline{\boldsymbol{A_{12}}} \cdot \boldsymbol{E_{inc,2}} \right)}{\overline{\boldsymbol{I}} - \varepsilon_b^2 \, \alpha_1 \alpha_2 \, \overline{\boldsymbol{A_{12}}} \cdot \overline{\boldsymbol{A_{21}}}}$$
$$\boldsymbol{p_2} = \frac{\varepsilon_0 \varepsilon_b \alpha_2 \left( \boldsymbol{E_{inc,2}} - \varepsilon_b \alpha_1 \, \overline{\boldsymbol{A_{21}}} \cdot \boldsymbol{E_{inc,1}} \right)}{\overline{\boldsymbol{I}} - \varepsilon_b^2 \, \alpha_1 \alpha_2 \, \overline{\boldsymbol{A_{21}}} \cdot \overline{\boldsymbol{A_{12}}}} \qquad (S25)$$

where it must be noted that the fraction here corresponds to a right-matrix division. With the expressions for both dipole moments, the total electric field at any point in space can be calculated as

$$\boldsymbol{E}(\boldsymbol{r}) = \boldsymbol{E_{inc}}(\boldsymbol{r}) + \overline{\boldsymbol{A}}(\boldsymbol{r} - \boldsymbol{r_1}) \cdot \boldsymbol{p_1} + \overline{\boldsymbol{A}}(\boldsymbol{r} - \boldsymbol{r_2}) \cdot \boldsymbol{p_2} \qquad (S26)$$

where $\boldsymbol{r_1}$ and $\boldsymbol{r_2}$ correspond to the position vectors of the two dipoles and the corresponding magnetic field can be obtained as $\boldsymbol{B}(\boldsymbol{r}) = \frac{1}{i\omega} \boldsymbol{\nabla} \times \boldsymbol{E}(\boldsymbol{r})$. Once the total field is obtained, the optical force on dipole $\boldsymbol{p_i}$ can be calculated by invoking the Lorentz Force equation [12], [13], [15]:

$$\boldsymbol{F_i} = \Re\mathrm{e}(\boldsymbol{p_i} \cdot \boldsymbol{\nabla}_i) \Re\mathrm{e}(\boldsymbol{E_i}) + \Re\mathrm{e}\left( \frac{d\boldsymbol{p_i}}{dt} \right) \times \Re\mathrm{e}(\boldsymbol{B_i}) \qquad (S27)$$

which for harmonic fields, can be time-averaged as

$$\langle \boldsymbol{F_i} \rangle = \Re\mathrm{e}[(\boldsymbol{p_i^*} \cdot \boldsymbol{\nabla}_i) E_i + i\omega \boldsymbol{p_i^*} \times \boldsymbol{B_i}] \qquad (S28)$$

where * denotes the complex conjugate. For incident plane waves, this expression can be simplified to the sum of the contributions of the incident light on the individual dipoles plus an interaction term, both of which require calculating the field across all the simulation space (equation S17) and the computationally costly gradients in equation S19 [13]. However, the complex Bessel beam requires full calculation using equation S19.

Formally speaking, approximation of nanoparticles as dipoles with polarizabilities $\alpha_i$ derives from the quasi-static approximation for spheres. However, this approximation can be extended to ellipsoidal objects and other geometries. For gold nanorods, we follow the expression derived by Kuwata et al. [16]:

$$\alpha \approx \frac{V}{\left( \Gamma + \frac{\varepsilon_b}{\varepsilon - \varepsilon_b} \right) + A_x(\Gamma)(n_b x)^2 + B_x(\Gamma)(n_b x)^4 - i \frac{4\pi^2 n_b^3}{3} \frac{V}{\lambda_0^3}} \qquad (S29)$$

where $V$ is the rod volume, $\lambda_0$ is the incident wavelength and $x = \frac{\pi a}{\lambda_0}$ is the size parameter, with $a$ being the particle size. $A_x(\Gamma)$ and $B_x(\Gamma)$ are functions dependent on the geometrical depolarization factor $\Gamma$. For rods, this factor is given by

$$\Gamma_{rod} = \frac{\left[ (\xi - 1)^3 - 2 - (\xi - 2\xi - 1)\sqrt{\xi^2 - 2\xi + 2} \right]}{3(\xi - 1)^3} \qquad (S30)$$

where $\xi$ is the aspect ratio between the long and short dimensions of the rod. The functions $A_x(\Gamma)$ and $B_x(\Gamma)$ are then

$$A_x(\Gamma) = -0.4865\Gamma - 1.046\Gamma^2 + 0.8481\Gamma^3$$
$$B_x(\Gamma) = 0.019\Gamma + 0.1999\Gamma^2 + 0.6077\Gamma^3 \qquad (S31)$$

An important parameter to consider is interparticle distance, since for closely positioned dipoles electrostatic contributions to the induced fields, neglected under the dipolar approximation, become relevant. This effect can be accounted for by other methods, such as the use of Mie theory or more complex DDA approaches. To elucidate the impact that this may have on our calculations, we compare our CDA approach to full wave FDTD results in the case of a linearly $y$-polarized plane wave with intensity 0.4 W/μm$^2$ intensity plane wave incident on the analyzed dimers, with a separation distance of 100 nm. This comparison is displayed in **Figure S4a**, and it can be seen that the overall shape of both the optical force and the extinction cross-section (calculated using the extinction theorem as shown in [17]) is quite similar.

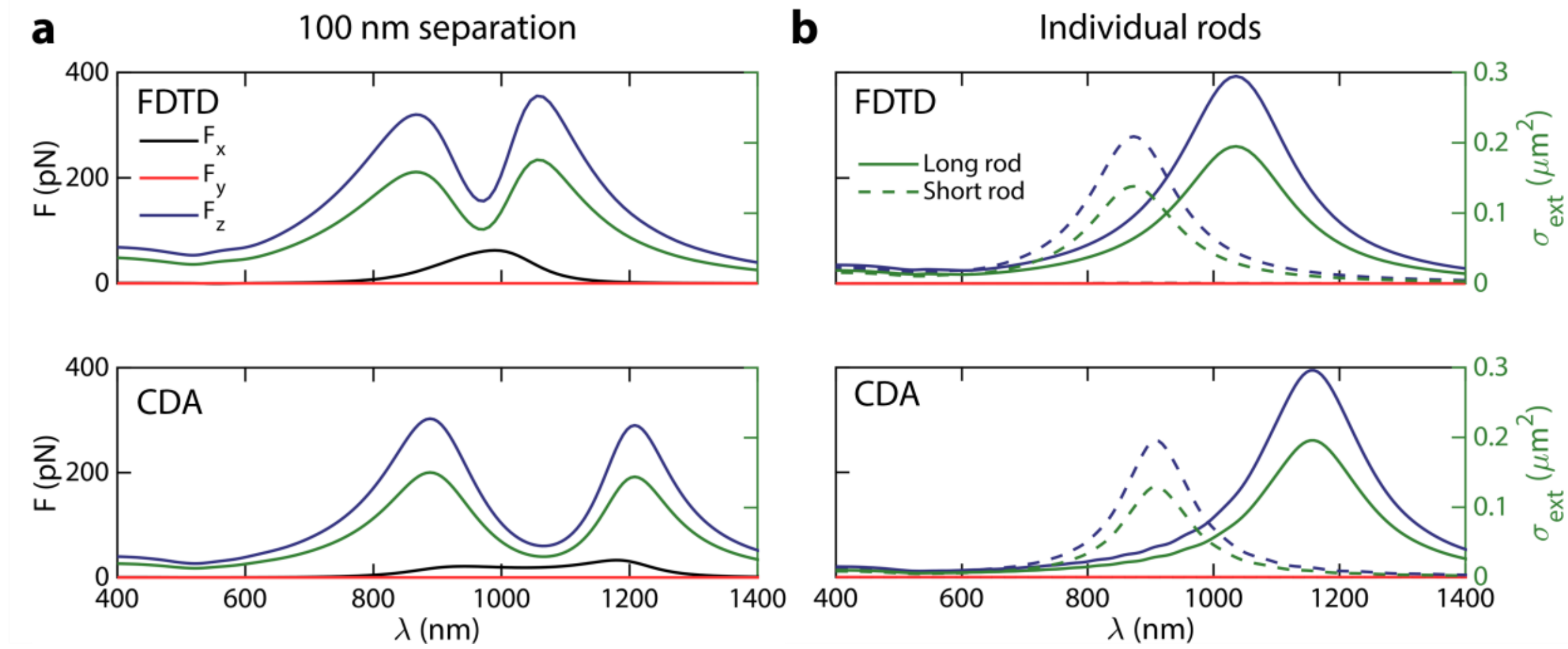

**Figure S4.** a) Comparison of FDTD and CDA solutions for a dimer with a 100 nm interparticle distance illuminated by a *y*-polarized plane wave with intensity 0.4 W/μm2, similarly to **Figure 1d** in the main text. b) FDTD and CDA results for the individual components of the dimer (distinguished by solid/dashed lines). The background refractive index for all calculations is set to $n_b = 1.5$.

The main difference between the CDA and FDTD methods is the location of the extinction peak corresponding to the larger rod in the dimer. While the FDTD peaks are located at 860 and 1050 nm, respectively; the CDA peaks are located at 880 and 1200 nm. The big redshift between the second peak locations between the methods mainly affects the longitudinal force, which closely follows the extinction cross-section, although the bigger gap between resonances in the CDA method flattens the transverse x-component force. Other than that, quantitative values do not change very significantly, with both longitudinal forces peaking around 300 pN.

While the interparticle distance might play a role in these discrepancies, the rod's polarizability modeling in equation S20 can highly influence the spectral locations of the individual resonances. **Figure S4b** shows the results of both methods for the individual rods. Here, the spectral positions of the peaks closely match the dimer results, with the CDA methods having a redshifted second peak with respect to FDTD simulations. Here, the FDTD peaks are located at 870 and 1035 nm, while the CDA peaks are at 905 and 1150 nm. This means that the polarizability modelling in equation S20 is the main responsible for the discrepancy of results, rather than the interparticle distance. It must be highlighted that the overall trends of forces and extinctions are similar between methods, meaning that the CDA method's precision is enough for our purpose in this work.

## 6. Transversal forces on plasmonic dimers

As seen in the previous sections, the optical force by an azimuthally polarized Bessel beam on a plasmonic nanoparticle will have a positive longitudinal component as well as a radial component. In a plasmonic dimer, the interaction between particles, driven by the coupling tensors $\overline{\boldsymbol{A}_{ij}}$ , might show important deviations from this behaviour. To check this, we provide in **Figure S5** the missing components from **Figure 2e-f**, $F_x$ and $F_y$. As in the CDA simulations the dimer was displaced from the center of the beam in the -x direction, $F_x = -F_\rho$ and $F_y = F_\varphi$ in the cylindrical coordinate system.

As expected, the longitudinal force, when the dimer is parallel with polarization, is accompanied by a strong x-component force, corresponding to a radial direction. In contrast, the y-component is near zero, as is expected of the azimuthal component for a single plasmonic particle.

On the other hand, when dimers are placed perpendicularly to the azimuthal polarization, as shown in the solid lines, all the components are near-zero, supporting the fact that independent control of the different degrees of freedom of the nanomotor can be achieved by orienting the dimers in this direction.

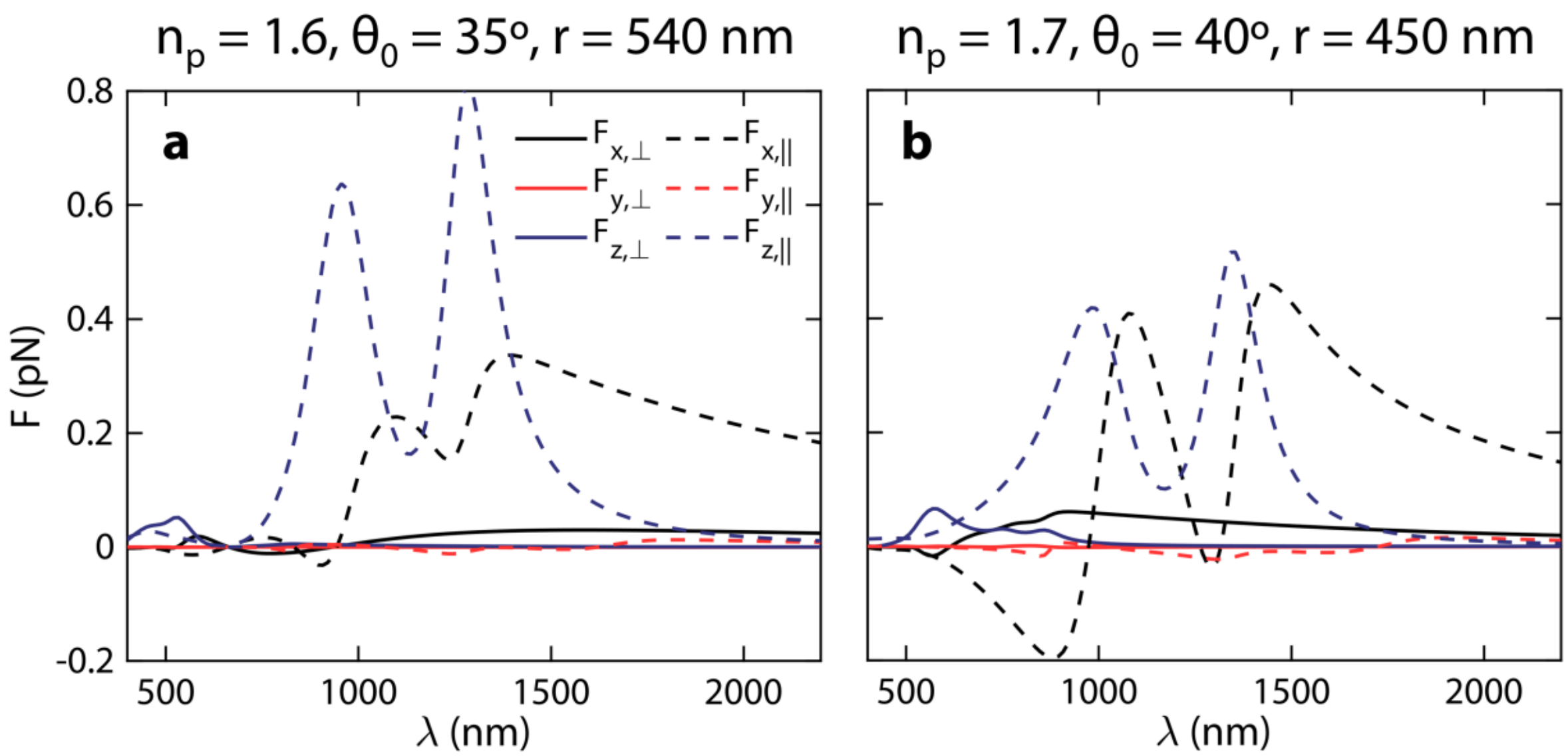

**Figure S5.** Optical forces on the highlighted cases from **Figure 2e** (a) and **Figure 2f** (b) with all components $F_x$, $F_y$ and $F_z$. Solid lines represent a perpendicular orientation to polarization, and dashed lines a parallel orientation with respect to polarization.

Some other relevant features can be found here. As shown in equation 4 in the main text, the longitudinal force is proportional to $\frac{\cos\theta_0}{\sin^2\theta_0}$, meaning that lower angles, as in **Figure S5a** with respect to **Figure S5b**, will generally have larger longitudinal forces. Regarding the radial force, its peaks are redshifted with respect to the longitudinal peaks, and the radial force has a remarkably weaker decay for longer wavelengths. This can be attributed to the fact that the radial force depends on the real part of the polarizability rather than the imaginary part. This prompts a different behavior, including the sign flip found at ~900 nm in **Figure S5b**.

Furthermore, in resonance, the force is positive (drawing the dimer to the center of the beam), which points to the dimer being located at a point where the product $J_1(x_\rho) \cdot J'_1(x_\rho)$ is negative. Since both functions are positive for small $x_\rho$, this suggests the opportunity to place the dimers closer to the center to avoid the radial force altogether. However, smaller distances between dimers might induce further coupling between elements, significantly altering the overall optical response.

## 7. Longitudinal forces and torques on nanomotors

For plane wave illumination, the longitudinal components of force and torque, not shown in **Figure 3** in the main text, are depicted in **Figure S6**. It can be seen again that the longer nanomotor, bigger in size and containing more dimers, is characterized by larger forces and torques. The $F_z$ component is much larger than either $F_{||}$ or $F_\perp$ in Figure 3, meaning that any plane wave illumination for lateral movement is going to be associated with a strong pushing force. Furthermore, the pushing force is significant for all considered wavelengths, which allows to control pushing movement independently of lateral movement if out-of-resonance wavelengths are employed (as, for example, $\lambda =$ 1700 nm). The longitudinal torque $T_z$, on the contrary, supposes a minor contribution against the $T_{||}$ component in **Figure 3**.

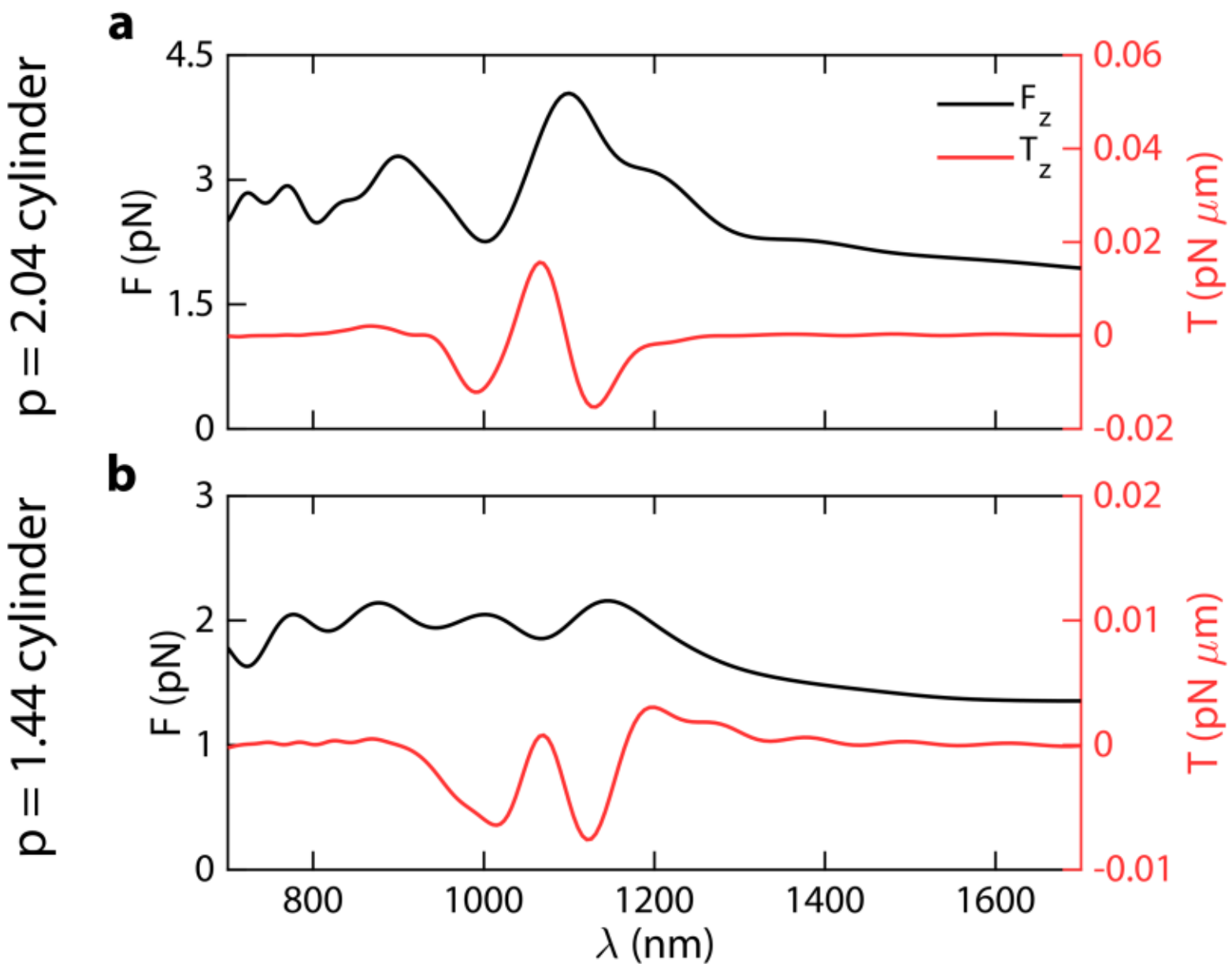

**Figure S6.** Longitudinal forces and torques $F_z$ and $T_z$ exerted by a linearly polarized plane wave with intensity 0.4 mW/ µm2 on the a) $p = 2.04$ and b) $p = 1.44$ aspect ratio nanomotors as a function of wavelength.

The longitudinal component torques for both nanomotors are shown in **Figure S7**. The magnitude of both torques is much lower than that of the dominant component $T_y$, especially in the case of the longer cylinder. This further allows the one-dimensional treatment employed in the diffusion simulations in the main text.

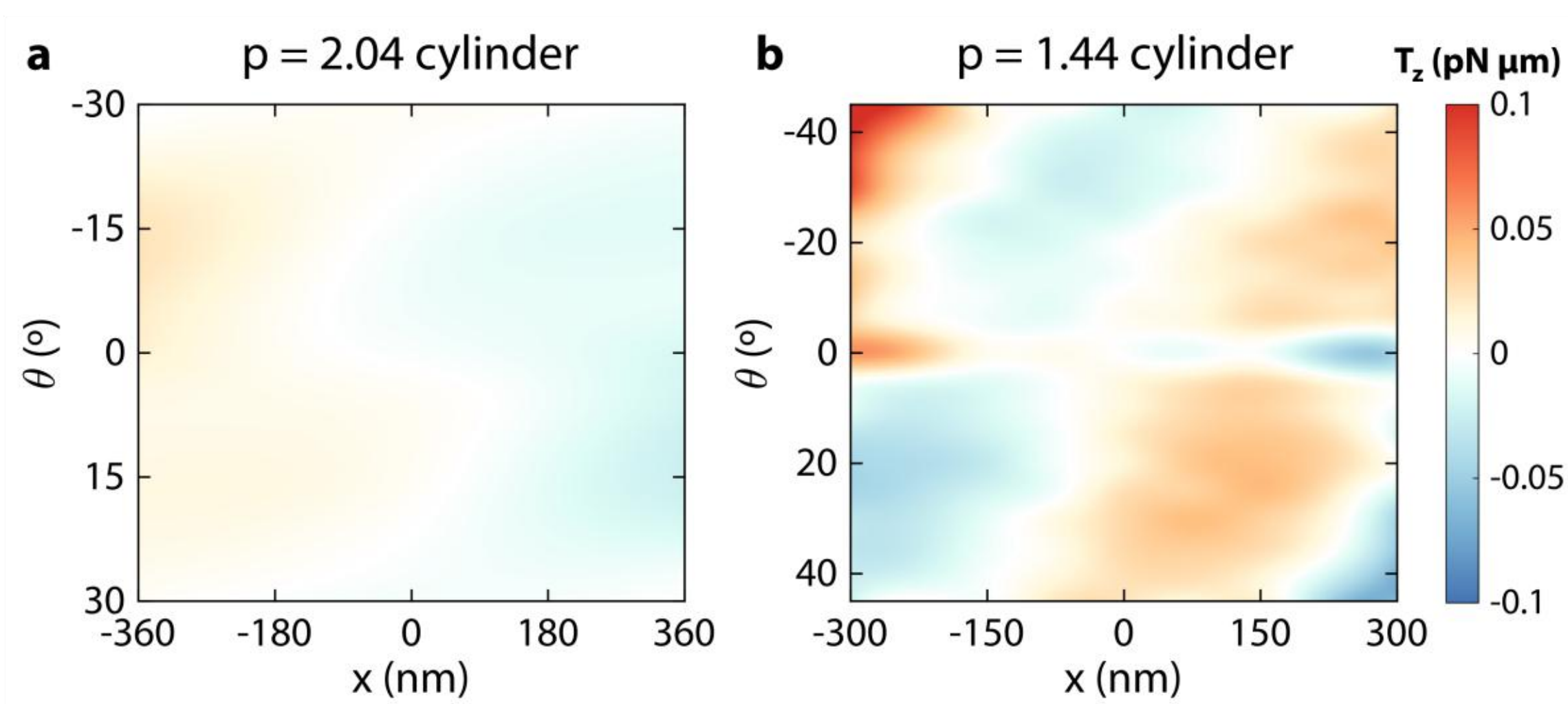

**Figure S7.** Longitudinal torque $T_z$ exerted on the a) $p = 2.04$ and b) $p = 1.44$ aspect ratio nanomotors for different positions and orientations.

## 8. Extension of the scheme for controlled rotations

The proposed scheme in the main text is designed to only provide translational lateral motion. However, for non-absorbing structures, a helicity-dependent optical torque $T_z$ can be exerted using circularly polarized light [18], [19]. Furthermore, as shown by Käll and coworkers [20], plasmonic nanorods can efficiently transduce the spin angular momentum from circularly polarized light into a rotatory motion, something that has been important to many different applications. In particular, plasmonic rods could be fitted into the design, providing a helicity-dependent rotatory motion that can be made independent from other lateral motions if said rods are tuned to a different illumination wavelength, i.e., a wavelength below 1000 nm, where scattering from the gold nanorod dimer becomes important.

Although absorption can play a role into the torque induced on such nanorods, the most important contribution comes from scattering processes, for which the time-averaged torque $T_z$ can be expressed as [21]

$$T_z \propto \Delta\alpha^2 \qquad (S32)$$

where $\Delta\alpha = \alpha_x - \alpha_y$ the polarizability anisotropy of the nanorods. As seen in this equation, longer rods, characterized by larger anisotropies, will allow higher torques. On the other hand, larger aspect ratios in plasmonic nanorods lead to a redshift of their plasmonic resonance [22]. We account for the necessary torque that these nanomotors require to overcome Brownian motion:

$$T_B = \sqrt{2\gamma_r k_B T} = \sqrt{k_B T \frac{64}{3} \pi \eta_W a^3} \qquad (S33)$$

where $\gamma_r = \frac{32}{3}\pi\eta_W a^3$ is the rotational friction coefficient for a cylinder with radius $a$ [23]. For the $p = 1.44$ design, the Brownian torque threshold is approximately 0.02 pN µm, which is relatively low (although higher than the $T_z$ achieved with linearly polarized illumination). Thus, rods with relatively low anisotropies can be employed for this purpose. To further separate the response of these rods from the translational dimers, we employ silver, whose resonances are typically found at shorter wavelengths [22], as the material of choice for these rods.

The proposed design is illustrated in **Figure S8a**. Silver nanorods (100 nm long, 50 nm wide and thick) are inserted between the spaces left by the four gold nanorod dimers. Their radial position is the same as the gold nanorod dimers (450 nm from the center of the $p = 1.44$ cylinder) and their orientation is perpendicular to the azimuthal orientation of the Bessel beam to preserve optical pulling. For the $p = 1.44$ cylinder nanomotor and appropriate Bessel beam illumination (40º, 1300 nm wavelegth), the forces and torques suffer almost negligible changes upon introduction of such silver nanorods, as demonstrated in **Figure S8b**. In particular, the optical pulling force is reduced by 5% approximately.

**Figure S8c-d** shows the transversal forces exerted on the $p = 1.44$ nanomotor by a circularly polarized wave, with and without the silver nanorods. Transversal forces have a similar pattern to their linearly polarized counterparts in **Figure 3** of the main text. However, as power is distributed between both x and y polarizations, both $F_x$ and $F_y$ have significant contributions above zero, although lower than their linearly polarized counterparts. This does not change much upon introduction of the silver nanorods, whose resonances are at shorter wavelengths. It is important to highlight that helicity changes just produce a permutation of the forces, as the nanomotor is not geometrically chiral.

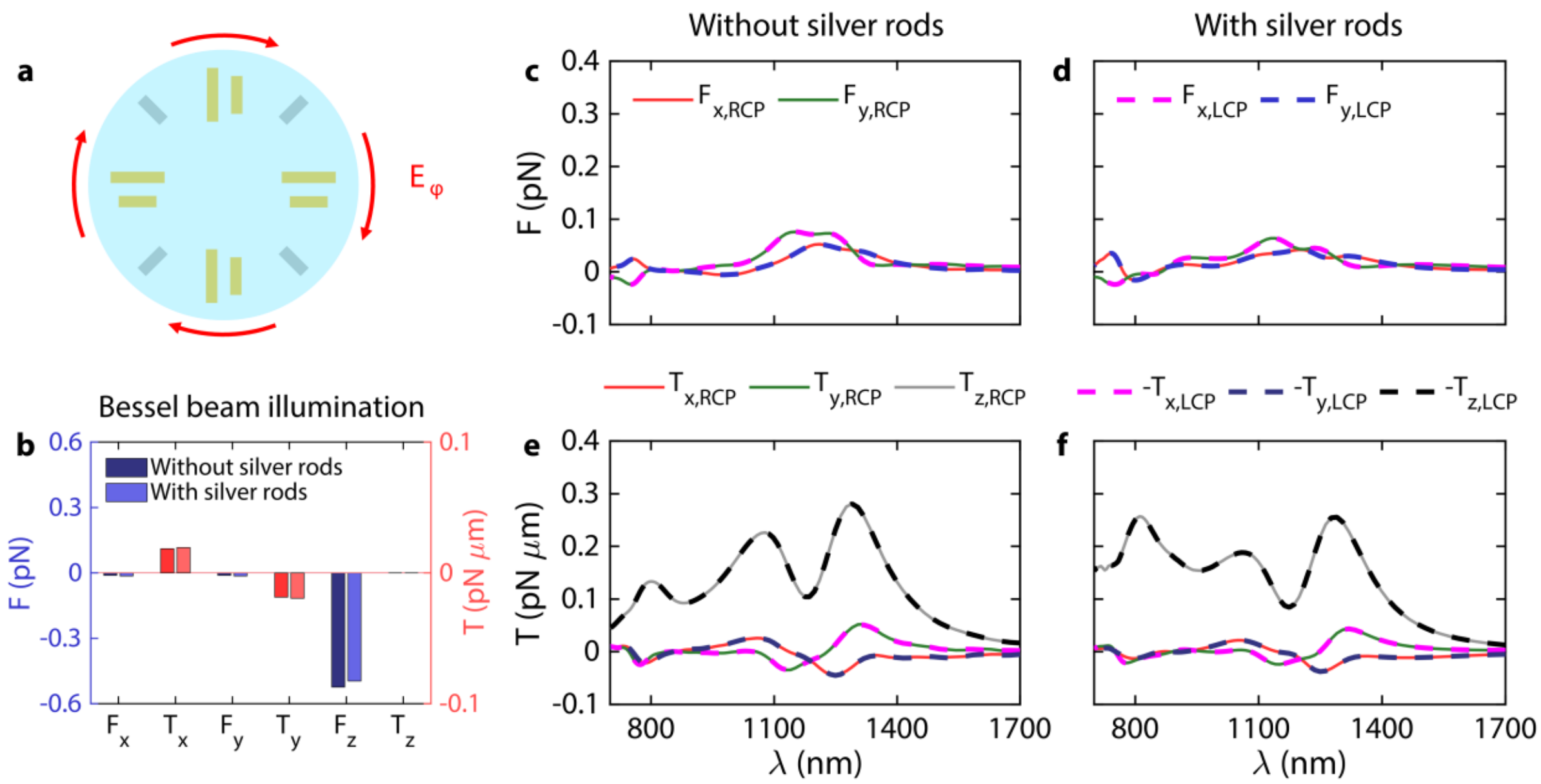

**Figure S8.** a) Scheme of the 3D nanomotor design, to which four silver nanorods have been added in the available space between the dimers (the center of the rods, as the dimers, is 450 nm away from the cylinder center). As with the dimers, the silver nanorods are perpendicularly oriented to the azimuthal polarization of a pulling Bessel beam. b) Comparison between the different force and torque components between the $p = 1.44$ nanomotor without and with the silver nanorods for the corresponding Bessel beam illumination. c) Optical transversal forces exerted by a 0.4 mW µm-2 circularly polarized plane wave on the $p = 1.44$ nanomotor without silver rods and (d) with them. Solid and dashed lines represent the different helicity (RCP or LCP) of the incoming wave. e) Optical torque exerted by the same circularly polarized plane wave on the $p = 1.44$ nanomotor without silver rods and (f) with them. Note that the LCP torque components, represented by dashed lines, have undergone a sign change to facilitate comparison with their RCP counterparts.

On the other hand, the torques, shown in **Figure S8e-f** have a very dominant $T_z$ component, which was not found for linear polarization (as shown in **Figure S6b**). Two peaks are found at 110 and 1300 nm, following the two resonances of the gold nanorods. These, however, are coincident with peak values of $F_x$, $F_y$, $T_x$ and $T_y$, meaning that $T_z$ cannot be independently controlled with these resonances. A third, weaker $T_z$ peak appears for a 800 nm wavelength, which is not correlated with strong forces or torques, and thus, usable for independent rotation control. This peak can be associated with the small lateral forces found at the same wavelength here and in **Figure S5b**.

Upon introduction of the silver nanorods, their contribution significantly enhances this weaker $T_z$ peak (around 92%), without significantly altering any of the other transversal forces and torques. Again, as the system is not chiral, the torques do not change in magnitude upon helicity changes. However, the transversal components are again interchanged, and all torques components undergo a sign flip. Thus, a helicity-dependent optical torque one order of magnitude above the Brownian threshold can be applied independently from other forces and torques, allowing efficient motion control over clockwise and anticlockwise rotations.

## 9. Characterization of the hydrodynamic properties of cylinders

The hydrodynamic behavior of cylinders has been the subject of intense research, following its frequent appearance in colloids, such as the tobacco mosaic virus or short DNA fragments [24], [25], [26]. Most works trying to describe these hydrodynamic properties have focused on long aspect-ratio ("rod-like") cylinders [27]. However, the aspect ratios of the cylinders described in this work are rather short, with aspect ratios $p \lesssim 2$. Therefore, the more general formulae developed by Ortega and García de la Torre are followed [28]. In this work, the diffusion tensor component $D_r^{\perp}$ is related to a rotational time $\tau_a$ by

$$\tau_a = \frac{1}{6D_r^{\perp}} \tag{S34}$$

For aspect ratios $p > 0.75$, as is the case in this work, $\tau_a$ can be calculated from the interpolating equation

$$\frac{\tau_a}{\tau_0} = 1.18 + 1.116(\ln p + 0.2877)^2 - 0.2417(\ln p + 0.2877)^3 + 0.4954(\ln p + 0.2877)^4 \tag{S35}$$

where $p = L/D$ is the length-to-diameter aspect ratio and $\tau_0 = \frac{\pi L^3 \eta_W}{4p^2 k_B T}$. Here, $\eta_W = 10^{-3}$ Pa·s is the water viscosity, $k_B$ is Boltzmann constant and $T$ is the temperature (taken as 300 K). On the other hand, the translational diffusion coefficient $D_t$ is given by

$$D_t = \frac{k_B T}{f_t} \tag{S36}$$

related to a translational friction coefficient $f_t$, given by a similar expression:

$$\frac{f_t}{f_{t0}} = 1.009 + 1.395 \cdot 10^{-2}(\ln p) + 7.880 \cdot 10^{-2}(\ln p)^2 + 6.040 \cdot 10^{-3}(\ln p)^3 \tag{S37}$$

with $f_{t0} = 6\pi\eta_W L \left(\frac{3}{16p^2}\right)^{1/3}$.